\documentclass[12pt]{article}
\usepackage{amsmath,color,empheq}
\usepackage[dvips]{graphicx}
\usepackage{amssymb}
\def\red#1{{\color{red} #1}}
\begin{document}
\def\prg#1{\par\medskip\noindent{\bf #1}}
\def\prgb#1{\par\noindent{\bf #1}}
\newcounter{nbr}\setcounter{nbr}{0}
\def\sitem#1{\addtocounter{nbr}{1}\item[(s\thenbr)] #1}
\def\eff{\,[=]\,}   

\def\lra{\leftrightarrow}              \def\Lra{{\Leftrightarrow}} \def\Ra{\Rightarrow}                   \def\La{\Leftarrow}
\def\nin{\noindent}                    \def\pd{\partial}
\def\dis{\displaystyle}                \def\vsm{\vspace{-8pt}}
\def\bull{\raise.25ex\hbox{\vrule height.8ex width.8ex}}
\def\ric{{Ric}}                       \def\ra{\rightarrow}
\def\Lie{{\cal L}\hspace{-.7em}\raise.25ex\hbox{--}\hspace{.2em}}
\def\hd{{^\star}}                      \def\mb#1{\hbox{{\boldmath $#1$}}}
\def\mbr#1{\hbox{{\red{\boldmath $#1$}}}}
\def\ph#1{\phantom{#1}}    \def\phb{\phantom{\Big|}}\def\diag{\text{diag}}
\def\det{\,\text{det}{}}
\def\tgr{GR$_{\parallel}$}             \def\mT{\hbox{$\mathbb{T}$}}

\def\hook{\hbox{\vrule height0pt width4pt depth0.3pt
\vrule height7pt width0.3pt depth0.3pt
\vrule height0pt width2pt depth0pt}\hspace{0.8pt}}
\def\inn{\hook}

\def\G{\Gamma}        \def\S{\Sigma}         \def\D{\Delta}
\def\a{\alpha}        \def\b{\beta}          \def\g{\gamma}
\def\d{\delta}        \def\m{\mu}            \def\n{\nu}
\def\th{\theta}       \def\k{\kappa}         \def\l{\lambda}
\def\vphi{\varphi}    \def\ve{\varepsilon}   \def\p{\pi}
\def\r{\rho}          \def\Om{\Omega}        \def\om{\omega}
\def\s{\sigma}        \def\t{\tau}           \def\eps{\epsilon}
\def\nab{\nabla}      \def\vth{\vartheta}

\def\cL{{\cal L}}     \def\bcL{\bar{\cL}}    \def\hu{\hat{u}}            \def\tR{{\tilde R}}   \def\cE{{\cal E}}      \def\hC{\hat{C}}
\def\cH{{\cal H}}     \def\hcH{\hat{\cH}}    \def\cT{{\cal T}}
\def\cK{{\cal K}}     \def\hcK{\hat{\cK}}    \def\cA{{\cal A}}
\def\cV{{\cal V}}     \def\tom{{\tilde\om}}  \def\cR{{\cal R}}
\def\bG{{\bar G}}     \def\wcH{{\check\cH}}  \def\wH{{\check H}}
\def\tcL{{\tilde\cL}} \def\tcH{{\tilde\cH}}  \def\hF{\hat{F}}
\def\orth{{\perp}}    \def\bm{{\bar m}}      \def\bn{{\bar n}}
\def\bi{{\bar\imath}} \def\bj{{\bar\jmath}}  \def\bk{{\bar k}}
\def\chm{\checkmark}  \def\bR{{\bar R}}      \def\chmr{\red{\chm}}
\def\bl{{\bar l}}     \def\br{{\bar r}}      \def\bs{{\bar s}}
\def\bp{{\bar p}}
\def\hs#1{\hspace{#1pt}}  \def\ul#1{\underline{#1}}
\def\ol#1{{\bar #1}}  \def\ir#1{\,{}^{#1}\hspace{-0.2pt}}
\let\Pi\varPi         \def\hpi{{\hat\pi}}   \def\hPi{{\hat\Pi}}
\def\bV{{\bar V}}     \def\bcH{{\bar\cH}}   \def\bT{{\bar T}}  \def\pfc{\text{PFC}}   \def\ub#1{\underbrace{#1}} \def\hu{{\hat u}}
\def\heps{\,{\hat\eps}} \def\hH{{\hat H}}   \def\bu{{\bar u}}
\def\bH{{\bar H}}      \def\fT{f(\mT)}      \def\chZ{{\check Z}}
\vfuzz=2.5pt 
\def\nn{\nonumber}
\def\be{\begin{equation}}             \def\ee{\end{equation}}
\def\ba#1{\begin{array}{#1}}          \def\ea{\end{array}}
\def\bea{\begin{eqnarray}}            \def\eea{\end{eqnarray} }
\def\beann{\begin{eqnarray*}}         \def\eeann{\end{eqnarray*} }
\def\beal{\begin{eqalign}}            \def\eeal{\end{eqalign}}
\def\lab#1{\label{eq:#1}}             \def\eq#1{(\ref{eq:#1})}
\def\bsubeq{\begin{subequations}}     \def\esubeq{\end{subequations}}
\def\bitem{\begin{itemize}}           \def\eitem{\end{itemize}}
\renewcommand{\theequation}{\thesection.\arabic{equation}}
\def\ns#1{{\normalsize #1}}  

\title{Local symmetries and physical degrees of freedom\\ in $\fT$ gravity: A Dirac Hamiltonian constraint analysis}

\author{Milutin Blagojevi\'c\footnote{\texttt{mb@ipb.ac.rs}}\\
\ns{Institute of Physics, University of Belgrade,
                      Pregrevica 118, 11080 Belgrade, Serbia}\\[5pt]
James M. Nester\footnote{\texttt{nester@phy.ncu.edu.tw}}\\
\ns{Department of Physics, National Central University,
    Chungli 32001, Taiwan},\\[-3pt]
\ns{Graduate Institute of Astronomy, National Central University,
    Chungli 32001, Taiwan}\\[-3pt]
\ns{and Leung Center for Cosmology and Particle Astrophysics}, \\[-3pt]
\ns{National Taiwan University,Taipei 10617, Taiwan} }

\date{}
\maketitle
\vspace{-3pt}

\begin{abstract}
In the literature on $\fT$ gravity, the status of local Lorentz invariance and the number of physical degrees of freedom have been controversial issues. Relying on a detailed Hamiltonian analysis, we show that there are several scenarios describing how local Lorentz invariance can be broken, but in the generic case, the number of physical degrees of freedom is found to be $N^*=5$; in $D$ dimensions, this number is $N^*=D(D-3)/2+(D-1)$. As expected, the theory is vulnerable to having problematical propagating modes. We compare our results with those existing in he literature. As a by-product of our analysis, the diffeomorphism invariance is explicitly confirmed.
\end{abstract}
\section{Introduction}
\setcounter{equation}{0}

The teleparallel theory of gravity (TG) can be understood as a gauge theory of local translations, with torsion as the only field strength. In the context of Poincar\'e gauge theory (PG) \cite{PG1,PG2}, a gauge theory of gravity with two field strengths, the curvature and the torsion, TG is naturally defined by the condition of vanishing curvature \cite{TG1}. In contrast to general relativity (GR), where the geometry of spacetime is characterized by a Riemannian curvature and vanishing torsion, TG has a nontrivial torsion but vanishing curvature. In spite of this \emph{geometric} difference, there is a special version of TG which is \emph{dynamically} equivalent to GR, known as the teleparallel equivalent of GR (TEGR or \tgr) \cite{TEGR1,TEGR2,TEGR3}. This fact is of particular importance for the physical interpretation of TG.

Experimental predictions of general relativity (GR) in the low energy limit (the Solar System), as well as in some high energy regimes, such as gravitational waves, have been extremely well tested \cite{will}. The situation with the observational data on the largest, cosmological scale is rather different. The standard cosmological model can explain most of these observations, but at the expense of introducing mysterious concepts of dark matter and dark energy, ``inferred to exist \emph{only} through their presumed gravitational effects" \cite{PG2}.

As a response to this challenging situation, there has been a lot of activity in developing modified (Riemannian and non-Riemannian) gravitational models; see, for instance, Refs. \cite{mod,fR}. One of the well-known models of this type is $f(R)$ gravity, which is defined as an extension of the GR Lagrangian, $\cL_R=-a_0 R$, to a function of $R$, $\cL_{fR}:=f(R)$. The existence of \tgr\ was a natural theoretical motivation to introduce an analogous extension of TG, known as $\fT$ gravity \cite{fT}. Due to the complicated  dynamical structure of $\fT$ gravity, its basic dynamical properties, such as the status of local Lorentz invariance and the number of physical degrees of freedom (d.o.f.), are still controversial; see Li at al. \cite{LMM11} and Ferraro and Guzm\'an \cite{Ferraro:2018tpu,Ferraro:2018axk}. The objective of the present work is to find out reliable answers to these controversies by a detailed analysis of $\fT$ gravity, based on Dirac's Hamiltonian approach \cite{Dirac64}.

This paper is organized as follows. In Section 2, we give a short account of TG, including the special case of \tgr, and describe its generalization to $\fT$ gravity. In Section 3, we use Dirac's Hamiltonian approach to examine the  canonical structure of the model. In particular, we found the preservation condition for the Lorentz constraint $C_{ij}$, which plays a central role in the canonical analysis of $\fT$ gravity. In Section 4, the Legendre transform technique is used to derive the Poisson bracket algebra between the Arnowitt-Deser-Misner (ADM) components of the Hamiltonian. Then, in Section 5, we continue by constructing the canonical generator of local translations, which allows us to prove the first-class nature of the Hamiltonians. In Section 6,  the preservation conditions of the Lorentz primary constraints are shown to produce a number of conditions on the corresponding multipliers. In the generic case, Lorentz invariance is completely broken and the number of d.o.f. is found to be $N^*=5$. In another interesting scenario, we obtained $N^*=2$. In Section 7, our analysis is compared to the works of Li et al. \cite{LMM11} and Ferraro and Guzm\'an \cite{Ferraro:2018tpu,Ferraro:2018axk}. Finally, we have a number of appendices which contain not only technical details, but also some interesting extensions of the main text. In particular, a short account of an alternative Hamiltonian analysis of $\fT$ gravity is given in Appendix \ref{appD}, and Appendix \ref{appF} generalizes the results of the main text to higher-dimensional spacetimes.

Our conventions are as follows. The latin indices $(i,j,\dots)$ are the local Lorentz indices, the greek indices $(\m,\n,\dots)$ are the coordinate indices, and both run over $0,1,2,3$; the orthonormal frame (tetrad) is
$\vth^i=\vth^i{}_\m dx^\m$ (1-form), $\vth=\det(\vth^i{}_\m)$, the dual basis (frame) is $e_i=e_i{}^\m\pd_\m$, the metric components in the local Lorentz and coordinate basis are $g_{ij}=(1,-1,-1,-1)$ and $g_{\m\n}=g_{ij}\vth^i{}_\m \vth^j{}_\n$, respectively, and $\ve_{ijmn}$ is the totally antisymmetric
symbol with $\ve_{0123}=1$.

\section{From teleparallel to \mb{\fT} gravity}\label{sec2}
\setcounter{equation}{0}

Poincar\'e gauge theory (PG) was developed in the early 1960s by localizing the Poincar\'e group of rigid symmetries of matter Lagrangians in Minkowski spacetime \cite{PG1,PG2}. The basic gravitational variables of PG are the tetrad field $\vth^i$ and the Lorenz connection $\om^{ij}$ (1-forms); the corresponding field strengths are the torsion $T^i:=d \vth^i+\om^i{}_j \vth^j$ and the curvature $R^{ij}:=d\om^{ij}+\om^i{}_k\om^{kj}$ (2-forms), and the underlying structure of spacetime is described by Riemann-Cartan geometry. In this framework, TG is defined by the condition of \emph{vanishing curvature}, $R^{ij}=0$ \cite{TG1}, which means that the related Lorentz connection is pure gauge. (For a more general approach based on metric-affine gravity, see Ref.~\cite{Obukhov2003}.) As a consequence, parallel transport is path independent (under certain topological restrictions on spacetime), and we have a teleparallel geometry, a  geometry with a \emph{distant (or absolute) parallelism}.

It seems quite natural to define the TG dynamics by a Lagrangian which is quadratic in the torsion field strength, but in that case, one has to keep in mind that the nonvanishing Lorentz connection is pure gauge. An efficient way to incorporate this information into the Lagrangian formalism is by imposing the condition $R^{ij}=0$ via a Lagrange multiplier method; see, for instance, \cite{TEGR2,Obukhov2003,mb-mv,Nester2017}. Such an approach is Lorentz-covariant and may be very useful in exploring the general structure of the theory, but it turns out to be rather complicated in certain practical calculations, as can be seen in the Hamiltonian formulations of \cite{TEGR2,mb-mv}. A significant technical simplification can be achieved by imposing the relation $\om^{ij}=0$ as the gauge-fixing condition. In the resulting dynamical model, usually called the \emph{(pure) tetrad} formulation of TG, the tetrad field remains the only dynamical variable and the torsion 2-form takes the simple form $T^i=d\vth^i$.

Is the tetrad form of TG equivalent to the Lorentz-covariant form?
A detailed analysis of this issue can be found in Ref.~\cite{Nester2017}, which also includes the case of $\fT$ gravity (for the case of \tgr, see Ref.~\cite{Obukhov2019}). Using the Lagrange multiplier approach, the authors were able to show that the related equations of motion contain the same information on the physical properties of the theory as the equations of the pure tetrad version. More general arguments supporting these results can be obtained from the Dirac canonical approach to gauge theories \cite{Dirac64}.
The Hamiltonian constraint analysis implies that the gauge-fixing conditions, by construction, merely remove unphysical (arbitrary) variables from the theory without affecting its observable (gauge-invariant) properties. Thus, one can fix the local Lorentz symmetry by taking the gauge condition $\omega^{ij}=0$.  One is then left with a purely frame theory which contains all the physics. In particular, this conclusion justifies using the pure tetrad approach to count the number of d.o.f. in $\fT$ gravity, or, for instance, to evaluate the conserved charges or entropy in \tgr\ \cite{jn-mbc}.

In our further analysis of the teleparallel theories, we rely, as did \cite{LMM11,Ferraro:2018tpu}, on the pure tetrad formalism, characterized by the vanishing spin connection. The general (parity even) TG Lagrangian is defined in terms of three independent quadratic invariants,
\be
L_{TG}=\vth\cL_{TG}\,,\qquad
       \cL_{TG}:=a_0T^{ijk}(h_1T_{ijk}+h_2T_{jik}+h_3g_{ij}T^m{}_{mk})\,,
\ee
where $a_0=1/16\pi G$. For the special choice of parameters
$(h_1,h_2,h_3)=\big(1/4,1/2,-1\big)$, one obtains the \tgr\ Lagrangian \cite{TEGR1,TEGR2,TEGR3}
\be
L_T=\vth\cL_T\,,\qquad
\cL_T=\frac{1}{4}a_0T^{ijk}(T_{ijk}+2T_{jik}-4g_{ij}V_k)\,,       \lab{2.2}
\ee
with $V_k:=T^m{}_{mk}$. The corresponding covariant momentum,
\be
H_{ijk}:=\frac{\pd L_T}{\pd T^{ijk}}=\vth\cH_{ijk}\, ,\qquad
\cH_{ijk}=a_0\big(T_{ijk}+2T_{[kj]i}-4g_{i[j}V_{k]}\big)\,,       \lab{2.3}
\ee
plays an important (technical) role in the canonical analysis of \tgr.
Variation of $L_T$ with respect to $\vth^i{}_\n$ yields the gravitational field equations, which turn out to coincide with the GR field equations in vacuum.

In the last two decades, in an attempt to find a physically acceptable description of the cosmological dynamics, many alternative gravitational models, based either on suitable modifications of GR or its non-Riemannian extensions,  have been proposed \cite{mod,fR,fT}. In particular, the form of $f(R)$ gravity motivates one to introduce an analogous Lagrangian for $\fT$ gravity,
\be
\cL_{fT}:=\fT\,, \qquad
\mT:=\cL_T\equiv
            \frac{1}{4}a_0T^{ijk}(T_{ijk}+2T_{jik}-4g_{ij}V_k)\,.\lab{2.4}
\ee
where $\mT$ is the the teleparallel counterpart of the Riemannian scalar $R$.

The Hamiltonian analysis of the covariant teleparallel gravity worked out in
\cite{TEGR2,mb-mv} shows that TG always has gauge generators associated with the local Poincar\'e symmetry of the underlying Riemann-Cartan spacetime: the local translations and the Lorentz symmetry of the frame-connection pair $(\vth^i,\om^{ij})$.  In addition, for the one special case of GR$_\|$, there is an additional local Lorentz symmetry that acts on the frame alone.  This \emph{pure frame local Lorentz symmetry} is absent for any other choice of the teleparallel Lagrangian.

After fixing the gauge $\om^{ij}=0$, the local Lorentz symmetry of the frame-connection pair is broken. The gauge symmetries of the generic tetrad form of TG are only local translations, whereas in the special case of GR$_\|$,
the pure frame Lorentz symmetry is not affected, it survives as a valid gauge symmetry of the theory. However, the situation in $f(\mT)$ gravity is more complicated. Although one might expect breaking of the pure frame local Lorentz symmetry, it is not a priori clear whether the violation is complete or only partial. As we shall see, the final answer depends on the complex dynamical structure of $\fT$ gravity.

In order to simplify the Hamiltonian analysis of $\fT$ gravity,
we find it convenient to represent $\fT$ as the Legendre transform of a function $V(\phi)$ \cite{fT,LMM11,Ferraro:2018tpu,Ferraro:2018axk}
\be
L^f=\vth\cL^f\, ,\qquad \cL^f:=\phi\mT-V(\phi)\, ,                 \lab{2.5}
\ee
where $\phi$ is an auxiliary scalar field. In classical mechanics, the Legendre transformation is a well-known technique used to switch from a Lagrangian $L(x,v)$ to the Hamiltonian $H(x,p)=pv-L(x,v)$. Based on an (incomplete) analogy with Brans-Dicke theory, the Lagrangian $\cL^f$ is often referred to as the scalar-tensor form of $\fT$ gravity. The new Lagrangian $\cL^f=\cL^f(\phi,\mT)$ is dynamically  equivalent to $\fT$. Indeed, using the relation $\mT=V'(\phi)$ obtained from $\d\cL^f/\d\phi=0$, one can express $\phi$ as a function of $\mT$, $\phi=\phi(\mT)$, provided $V'$ is invertible. Then, by substituting $\phi(\mT)$ into $\cL^f$, it becomes a function of $\mT$ only, which confirms the equivalence with $\fT$. The function $V'(\phi)$ is invertible in the domain where $V''(\phi)\ne 0$.

In what follows, our analysis of $\fT$ gravity will be based on the Lagrangian \eq{2.5}, with the basic dynamical variables $(\vth^i{_\m},\phi)$. The convenience of this formalism is clearly visible in the form of the covariant momentum
\be
H^f_{ijk}:=\frac{\pd L^f}{\pd T^{ijk}}=\phi H_{ijk}\, ,\qquad   \cH^f_{ijk}:=\phi\cH_{ijk}\, ,
\ee
which is obtained from the \tgr\ expression \eq{2.3} by the simple rule
$a_0\to a_0\phi$.

Variation of $L^f$ with respect to $\vth^i{_\m}$ and $\phi$ yields the field equations in vacuum,
\bsubeq
\bea
\cE_i{^\n}&:=&-\frac{\d L^f}{\d \vth^i{_\n}}
 =\nab_\m(\phi H_i{}^{\m\n})+T_{mni}(\phi H^{mn\n})-e_i{^\n}L^f    \nn\\
 &\equiv&(\pd_\m\phi)H_i{}^{\m\n}
    +\phi\Big[\nab_\m H_i{}^{\m\n}+T_{mni}H^{mn\n}-e_i{^\n}\vth\mT\Big]
    +e_i{^\n}\vth V(\phi)=0 \, ,                                     \\
\cE_\phi&:=&\vth\Big[\mT-\pd_\phi V(\phi)\Big]=0\, .           \lab{2.7b}
\eea
\esubeq
In the presence of matter, the right-hand sides contain the corresponding matter currents.

Using the identities from Appendix A of \cite{TEGR2}, the first equation can be transformed into
\bea
\cE^{ik}=-(\pd_\m\phi)H^{ik\m}-\phi\Big[2a_0\vth\Big(R^{ik}(\tom)
     -\frac{1}{2}g^{ik}R(\tom)\Big)\Big]+g^{ik}\vth V(\phi)=0\,,\lab{2.8}
\eea
where $\tom$ is the Riemannian connection. Its trace and antisymmetric part read
\bsubeq\lab{2.9}
\bea
&&\cE^k{_k}=4\vth\big[a_0V^\m\pd_\m\phi+a_0\phi\tR/2+V(\phi)\big]=0\,,\\
&&\cE^{[ik]}=-H^{[ik]\m}\pd_\m\phi=0\, .                         \lab{2.9b}
\eea
\esubeq
In \tgr, the six equations \eq{2.9b} are trivial.

In our further exposition, the superscripts `$f$' will be omitted, to simplify the notation.

\section{Hamiltonian form of \mb{\fT} gravity}\label{sec3}
\setcounter{equation}{0}

The Dirac Hamiltonian formalism for a system with constraints \cite{Dirac64} is a particularly suitable approach to analyze both local symmetries and the dynamical d.o.f. in $\fT$ gravity. As we shall see,  the analysis closely follows certain aspects of the \tgr\ structure at an early stage, but later, differences become more and more serious.

In the present analysis, we rely on the $(1+3)$ decomposition of spacetime, whose basic aspects can be characterized by two simple properties: (p1) at each point of a spatial hypersurface $\S: x^0=$ const., one can define a unit timelike vector $\mb{n}=(n_k)$, orthogonal to $\S$; (p2) any spacetime vector $\mb{V}=(V_k)$ can be decomposed into a component $V_\orth:=n^k V_k$ along $\mb{n}$ and another component $V_\bk:=V_k-n_k V_\orth$ laying in the tangent space of (``parallel" to) $\S$. As a consequence, $n^k V_\bk=0$. We will use $\ve_{\bi\bj\bk}$ as the Euclidean epsilon symbol with
$\ve_{\bar 1\bar 2\bar 3}=1$, and $\d^{ijk}_{mnl}$ is the generalized Kronecker symbol.

\subsection{Primary constraints}

The canonical momenta $(\pi_i{^\m},\pi_\phi)$ associated to the basic Lagrangian variables $(\vth^i{_\m},\phi)$ are
\be
\pi_i{^\m}=\frac{\pd L}{\pd T^i{}_{0\m}}=\phi H_i{}^{0\m}\, ,
           \qquad \pi_\phi=\frac{\pd\cL}{\pd_0\phi}\,=0\,.         \lab{3.1}
\ee
Note first that these relations define the set of $4+1$ primary constraints
\be
\pi_i{^0}\approx 0\,,\qquad \pi_\phi\approx 0\,,
\ee
the existence of which does not depend on the particular values of the coupling constants. On the other hand, the canonical momentum $\pi_i{}^\a$ can be equivalently expressed in terms of the parallel canonical momentum
\bsubeq
\bea
&&\hpi_{i\bk}:=\pi_i{^\a}\vth_{k\a}=\phi J\cH_{i\orth\bk}\, ,    \lab{3.3a}\\
&&\cH_{i\orth\bk}=
  a_0\Big[T_{i\orth\bk}+(T_{\bk\orth i}-T_{\orth\bk i})
                   -2(n_i V_\bk-g_{i\bk}V_\orth)\Big]\, ,             \nn
\eea
where $J:=\vth/N$. Then, the constraint content of the first relation in \eq{3.1} can be clarified by introducing a new object $P_{i\bk}$,
\be
P_{i\bk}:=\hpi_{i\bk}/J-\phi\cH_{i\orth\bk}(0)=\phi\cH_{i\orth\bk}(1)\,,
\ee
\esubeq
where $\cH_{i\orth\bk}(0)$ does not depend on the ``velocities" $T_{i\orth\bk}$, and $\cH_{i\orth\bk}(1)$ is linear in them. The irreducible decomposition of $P_{i\bk}$ with respect to the group of three-dimensional spatial rotations yields \cite{TEGR2}
\bsubeq\lab{3.4}
\bea
&&P_{\orth\bk}\equiv
     \hpi_{\orth\bk}/J+2a_0\phi\,T^\bm{}_{\bm\bk}\approx 0\,,       \nn\\
&&\ir{A}P_{\bi\bk}\equiv \ir{A}\hpi_{\bi\bk}/J
    -a_0\phi\,T_{\orth\bi\bk}\approx 0\,,~                  \lab{3.4a}\\[5pt]
&&P^\bm{}_\bm\equiv\hpi^\bm{}_\bm/J=4a_0\phi\,T^\bm{}_{\bm\orth}\,,\nn\\
&&\ir{T}P_{\bi\bk}\equiv\ir{T}\hpi_{\bi\bk}/J
        =2a_0\phi\ir{T}T_{\bi\orth\bk}\, ,                       \lab{3.4b}
\eea
\esubeq
where $\ir{A}X_{\bi\bk}=X_{[\bi\bk]}$, $\ir{T}X_{\bi\bk}=X_{(\bi\bk)}-g_{\bi\bk}X^\bn{}_\bn/3$. Now, it is obvious that the first two relations define 6 additional primary constraints. Further calculations are greatly simplified by representing these primary constraints in a compact form:
\bsubeq\lab{3.5}
\be
C_{ik}=\cH_{ik}+a_0\phi B_{ik}\,,                                    \\
\ee
where
\bea
&&\cH_{ik}:=\hpi_{i\bk}-\hpi_{k\bi}=2\pi_{[i}{}^\a \vth_{k]\a}\, ,  \nn\\
&&B_{ik}:=\pd_\a B^{0\a}_{ik}\, ,\qquad
  B_{ik}^{0\a}\equiv\ve^{0\a\b\g}_{ikmn}\vth^m{_\b}\vth^n{_\g}\,,
\eea
\esubeq
and we used the notation $\ve^{0\a\b\g}_{ikmn}:=\ve^{0\a\b\g}\ve_{ikmn}$.
The above result follows from the identity $B_{ik}=-2J(T^\orth{}_{\bi\bk}-n_iT^\bm{}_{\bm\bk}+n_kT^\bm{}_{\bm\bi})$. The existence of the constraints $C_{ij}$ is caused by the special values of the coupling constants in the \tgr\ Lagrangian \eq{2.2}. On the other hand, the remaining two relations \eq{3.4b} are not constraints, they relate the velocities $T_{i\orth\bk}$ to the canonical momenta $\hpi_{i\bk}$.

\subsection{Hamiltonians}

Starting with the definition of the canonical Hamiltonian,
\be
H_c:=\pi_i{^\a}\pd_0 \vth^i{_\a}-b\cL\, ,
\ee
one can rewrite it in the standard Dirac-ADM form \cite{DADM},
\bsubeq\lab{3.7}
\be
H_c=N\cH_\orth+N^\b\cH_\b+\pd_\a D^\a\, ,
\ee
where
\bea
&&\cH_\orth:=\hpi_i{^\bm}T^i{}_{\orth\bm}-J\cL-n^i\pd_\a\pi_i{^\a}\,, \nn\\
&&\cH_\b:=\pi_i{^\a}T^i{}_{\b\a}-\vth^i{_\b}\pd_\a\pi_i{^\a}\, ,      \nn\\
&&D^\a:=\pi_i{^\a}\vth^i{_0}\, .                                   \lab{3.7b}
\eea
\esubeq
Here, the lapse and shift functions $N=n_k \vth^k{_0}$ and $N^\a=e_\bk{^\a}\vth^k{_0}$, respectively, are linear in the unphysical variables $\vth^k{_0}$. The lapse Hamiltonian $\cH_\orth$ is the only dynamical part  of $H_c$ as it depends on the Lagrangian. To eliminate the velocities $T_{i\orth\bk}$ from $\cH_\orth$, we use the relation
\bsubeq\lab{3.8}
\be
\hpi_i{^\bm}T^i{}_{\orth\bm}-J\cL=
                \frac{1}{2}JP^{i\bm}T_{i\orth\bm}-J\bar\cL\,,      \lab{3.8a}
\ee
where $\bar\cL:=\cL(0)\equiv\phi\bar\mT-V(\phi)$. Then, inserting here the irreducible decomposition
\be
P^{i\bm}T_{i\orth\bm}=\frac{1}{3}P^\bm{}_\bm T^{\bn}{}_{\orth\bn}
    +\ir{T}P_{\bi\bk}\ir{T}T^{\bi\orth\bk}
    +\Big[\ir{A}P_{\bi\bk}\ir{A}T^{\bi\orth\bk}
    +P_{\orth\bk}T^{\orth\orth\bk}\Big]\, ,                        \lab{3.8b}
\ee
\esubeq
where the last two terms are weakly vanishing as a consequence of \eq{3.4a}, the elimination of velocities with the help of \eq{3.4b} yields
\bea
&&\cH_\orth\approx \frac{1}{2a_0\phi}P^2
             -J(\phi\bar\mT-V)-n^i\pd_\a\pi_i{^\a}=:\wcH_\orth\,,    \nn\\
&&P^2:=\frac{1}{2J}\Big[\hpi_{(\bm\bn)}\hpi^{(\bm\bn)}
                           -\frac{1}{2}(\hpi^\bm{}_\bm)^2\Big]\,,    \nn\\
&&\bar\mT:=\frac{1}{4}a_0\Big(T_{i\bm\bn}T^{i\bm\bn}
                +2T_{\bi\bm\bn}T^{\bm\bi\bn}
          -4T^\bm{}_{\bm\bk}T_\bn{}^{\bn\bk}\Big)\, .              \lab{3.9}
\eea
An explicit distinction between $\cH_\orth$, defined by \eq{3.7b}, and its on shell version $\wcH_\orth$ is usually disregarded, but we keep it for later convenience. Apart from the $V$ term, the rest of $\wcH_\orth$ is obtained from the \tgr\ expression by $a_0\to a_0\phi$.

The general Hamiltonian dynamics is described by the total Hamiltonian
\be
H_T=\wH_c+u^i{_0}\pi_i{^0}+u_\phi\pi_\phi
                                +\frac{1}{2}u^{ij}C_{ij}\,,       \lab{3.10}
\ee
where $u's$ are, at this stage, arbitrary Hamiltonian multipliers. Their
dynamical interpretation is given by (Appendix \ref{appA})
\bsubeq\lab{3.11}
\bea
&&u_\phi=\pd_0\phi=N\pd_\orth\phi+N^\a\pd_\a\phi\, ,                 \\
&&u^{\orth\bn}=NT^{\orth\orth\bn}\, ,\qquad
   u^{\bm\bn}=N\ir{A}T^{\bm\orth\bn}\, .                          \lab{3.11b}
\eea
\esubeq

Keeping the last two terms in \eq{3.8b}, the complete
dynamical Hamiltonian $\cH_\orth$, defined in \eq{3.7b} as the Legendre transform of $\cL$ with respect to the velocity $T^i{}_{\orth\bm}$, is given by
\be
\cH_\orth=\wcH_\orth+\frac{1}{2}\hu^{mn}C_{mn}\, ,\qquad
                     \hu^{mn}:=N^{-1}u^{mn}\, .                   \lab{3.12}
\ee
Simultaneously, the expression for the total Hamiltonian is simplified,
\be
H_T=H_c+u^i{_0}\pi_i{^0}+u_\phi\pi_\phi\,.
\ee

\subsection{Preservation of primary constraints}

For consistency of the Hamiltonian analysis, every constraint $\vphi$ appearing in the theory has to be preserved during dynamical evolution of the system, determined by the total Hamiltonian as
\be
\chi:=\pd_0\vphi=\{\vphi,H_T\}\approx 0\, .                       \nn
\ee
In the expressions \eq{3.10} or \eq{3.12} for $H_T$, an integration over $d^3x$ is implicitly understood. Let us now apply this condition to the primary constraints $\vphi_A=(\pi_i{^0},\pi_\phi,C_{ij})$.

Starting with the preservation condition for $\pi_i{^0}$, one finds
\bsubeq
\bea
&&\chi_i:=-\{\pi_i{^0},H_T\}=n_i\cH_\orth+e_\bi{^\a}\cH_\a\approx 0,\\
&&\chi_\orth=\cH_\orth\approx 0\,,\qquad \chi_\a=\cH_\a\approx 0\,.
\eea
\esubeq

To calculate $\chi_\phi:=\pd_0\pi_\phi$, we use the relations
\bsubeq\lab{3.15}
\bea
&&\{\pi_\phi,\wcH_\orth\}=-\pd_\phi\wcH_\orth
   =\frac{1}{2a_0\phi^2}P^2+J(\bar\mT-\pd_\phi V)=:F_\phi\,,   \lab{3.15a}\\
&&\{\pi_\phi,C_{ij}\}=-\pd_\phi C_{ij} =-F_{ij}\, ,
  \qquad F_{ij}:=a_0B_{ij}\,,                                  \lab{3.15b}
\eea
\esubeq
where the $\d$ functions on the right-hand sides are omitted, which imply
\be
\chi_\phi=NF_\phi-\frac{1}{2}u^{mn}F_{mn}\approx 0\,.           \lab{3.16}
\ee
As shown in Appendix \ref{appB}, the Lagrangian counterpart of $\chi_\phi$ is $\vth(\mT-\pd_\phi V)$.

The key preservation condition at the level of primary constraints is that of the Lorentz constraint $C_{ij}$, which is of particular importance for a proper understanding of the status of Lorentz invariance in $\fT$ gravity. A direct calculation shows that the expression $\chi_{ij}:=\{C_{ij},H_T\}$ has the form (Appendix \ref{appC})
\bsubeq\lab{3.17}
\be
\chi_{ij}=G_{ij}{^k}(\pd_k\phi)\d\approx 0\, ,
\ee
where
\bea
&&G_{ij}{^\orth}:=-2a_0\vth(T^\orth{}_{\bi\bj}-n_i\bV_\bj+n_j\bV_\bi)
                =NF_{ij}\, ,                                      \\[2pt]
&&G_{ij}{}^\bk:=2a_0J\d^{\orth\bk\bn}_{ijm}u^m{_\bn}
  -N\Big[\phi^{-1}g^{\bk\bm}(n_i\hpi_{(\bm\bj)}-n_j\hpi_{(\bm\bi)})
         +a_0J\d^{\bk\bm\bn}_{\bi\bj\br}T^\br{}_{\bm\bn}\Big]\,.
\eea
\esubeq
The result is also verified in the Lagrangian approach by showing that the Hamiltonian transcription of the six Lagrangian equations \eq{2.9b} coincides with $\chi_{ij}$.

Based on the analysis presented in Appendix \ref{appC}, one can explicitly determine the Poisson bracket (PB) of $C_{ij}$ with itself, relying on the value of the coefficient $B_3$. The result can be conveniently written in the form
\be
u^{mn}\{C_{ij},C_{mn}\}=2u^{mn}(g_{jm}C_{in}-g_{im}C_{jn})
           +4a_0J\phi_{\bk}\d_{ijm}^{\orth\bk\bn}u^m{}_n\,.     \lab{3.18}
\ee
The presence of the nontrivial last term shows that at least some components of $C_{ij}$ are second class constraints, which means that local Lorentz invariance is at least partially broken. A detailed analysis of this issue appears later in section 6.

The preservation of the primary constraints $\vphi_A$ leads to the corresponding secondary constraints,
\be
\chi_A:=(\wcH_\orth,\cH_\a;\chi_\phi,\chi_{ij})\,.
\ee
Since $f(\mT)$ gravity is expected to be invariant under local translations (diffeomorphisms), the Hamiltonians $(\wcH_\orth,\cH_\a)$ (or their suitable deformations) are expected to be first class (FC). On the other hand, the constraints $(\chi_\phi,\chi_{ij})$ define seven conditions on the seven multipliers $(u_\phi,u^{mn})$. These conditions will play a prominent role in the forthcoming analysis of local Lorentz invariance, as well as in counting  the physical d.o.f..

For a comparison of the results found in the present analysis to the canonical structure based on the standard Lagrangian $\cL_{fT}=f(\mT)$, see Appendix \ref{appD}.

\section{PB algebra between Hamiltonians}\label{sec4}
\setcounter{equation}{0}

In this section, we wish to examine the PB algebra between the Hamiltonians  $(\cH_\orth,\cH_\a)$, which will allow us to understand their preservation conditions, as well as the status of diffeomorphisms invariance in $f(\mT)$ gravity.

In the generic TG, the Hamiltonians satisfy the same PB algebra as in GR,
\bsubeq\lab{4.1}
\bea
&&\{\cH_\a,\cH'_\b\}=(\cH_\b\pd_\a-\cH'_\a\pd'_\b)\d\,,         \lab{4.1a}\\
&&\{\cH_\a,\cH'_\orth\}=\cH_\orth\pd_\a\d\, ,                   \lab{4.1b}\\
&&\{\cH_\orth,\cH'_\orth\}\approx -\big(\ir{3}g^{\a\b}\cH_\b
       +\ir{3}g^{\prime\a\b}\cH'_\b\big)\pd_\a\d\,.             \lab{4.1c}
\eea
\esubeq
An elegant proof can be found in Mitri\'c \cite{PM19}. Instead of relying on explicit expressions for $(\cH_\a,\cH_\orth)$, he followed a more effective method introduced by Nikoli\'c \cite{IN92}, based on treating the dynamical Hamiltonian $\cH_\orth$ as a Legendre transform of the Lagrangian, see \eq{3.7b}.

In \tgr, the existence of parameters with critical values generates the additional constraints $C_{ij}$. In \cite{PM19}, the author used the Legendre transform approach to show that the PB algebra of the Hamiltonian constraints remains the same as in \eq{4.1}. Note that the explicit expressions for $\cH_\orth$ in TG and in \tgr\ are different.

The case of $f(\mT)$ gravity requires further generalizations. When $\cH_\a$ acts on the phase-space variables $(\vth^i{_\a},\pi_j{^\b})$ via the PB operation, it generates spatial diffeomorphisms. The  presence of  the additional phase-space variables $(\phi,\pi_\phi)$ suggests to modify $\cH_\a$ by adding the term $\pi_\phi\pd_\a\phi$. A consistent realization of this idea requires one to relocate the term $\pi_\phi\pd_0\phi$ from $H_T$ to $H_c$, whereby both $\cH_\a$ and $\cH_\orth$ are effectively modified as
\bsubeq\lab{4.2}
\bea
&&\bcH_\a:=\cH_\a+\pi_\phi\pd_\a\phi\,,                          \lab{4.2a}\\
&&\bcH_\orth:=\cH_\orth+\pi_\phi\pd_\orth\phi\,.                 \lab{4.2b}
\eea
\esubeq
Here, $\cH_\orth$ as understood as the Legendre transform of $\cL$,
\be
\cH_\orth=\hpi_i{^\bm}T^i{}_{\orth\bm}-J\cL-n^i\pd_\a\pi_i{^\a}\,.\lab{4.3}
\ee
With these modifications, the total Hamiltonian can be written in a more compact form as
\bsubeq\lab{4.4}
\bea
&&H_T=H_c+u^i{_0}\pi_i{^0}+\pd_\a D^\a\,,                          \\
&&H_c=\vth^i{_0}\bcH_i=N\bcH_\orth+N^\a\bcH_\a\,,
\eea
\esubeq
where the primary constraints $C_{ij}$ and $\pi_\phi$ are hidden inside $H_c$.

Now, we are ready  to find the PBs between the modified Hamiltonians \eq{4.2}. In the approach based on explicit expressions for the Hamiltonians, the most difficult part of the calculation stems from the fact that $\cH_\orth$ depends on the Lagrangian. Studying the more complicated case of PG, Nikoli\'c \cite{IN92} used a different strategy, based essentially, but not entirely, on treating the dynamical Hamiltonian as a Legendre transform of $\cL$ with respect to the velocities $T_{i\orth\bk}$, in accordance with \eq{4.3}. Applying certain identities characterizing Legendre transformations, he was able to derive the PB algebra of the Hamiltonian constraints without specifying the explicit form of the Lagrangian. Based on a number of technical details that can be found in \cite{PM19}, we use here an analogous approach which takes into account the presence of an extra dynamical variable $\phi$ in $f(\mT)$ gravity. As shown in Appendix \ref{appE}, the final result takes the form
\bsubeq\lab{4.5}
\bea
&&\{\bcH_\a,\bcH'_\b\}\approx
                       (\bcH_\b\pd_\a-\bcH'_\a\pd'_\b)\d\,,     \lab{4.5a}\\
&&\{\bcH_\a,\bcH'_\orth\}\approx\bcH'_\orth\pd_\a\d\,,          \lab{4.5b}\\
&&\{\bcH_\orth,\bcH'_\orth\}\approx -\big(\ir{3}g^{\a\b}\bcH_\b
       +\ir{3}g^{\prime\a\b}\bcH'_\b\big)\pd_\a\d\,.            \lab{4.5c}
\eea
\esubeq

Is the result \eq{4.5} sufficient to conclude that $\bcH_\a$ and $\bcH_\orth$ are FC? Certainly not, since there are other constraints in the theory, whose PBs with the Hamiltonians are still not known.
In order to extend the previous considerations, one can calculate the PBs between the Hamiltonian constraints and $C_{mn}$. Using the form of the term $B_2$ in Appendix \ref{appC}, Eq. \eq{3.15b}, and the relations \eq{4.4}, one can show that both $\bcH_\a$ and $\bcH_\orth$ have vanishing PBs with $C_{mn}$. However, that is still not sufficient since the PBs of the Hamiltonian constraints with $\chi_\phi,\chi_{ij}$ and the related, possibly nontrivial preservation conditions, are not yet known.

A refined analysis in the next section will allow us to go a step further.

\section{Diffeomorphism invariance}\label{sec5}
\setcounter{equation}{0}

In this section, we construct the Hamiltonian gauge generator for local translations in $f(\mT)$ gravity, based on Castellani's algorithm \cite{LC82}, and use it to show that $(\bcH_\orth,\bcH_\a)$ are FC.

If the local symmetries of a gauge theory are described only in terms of the gauge parameters $\xi^i(x)$ and their first derivatives, the canonical gauge generator has the form (integration over $d^3 x$ understood)
\be
G=\dot\xi^i G^0_i+\xi^i G^1_i\, ,
\ee
where the phase-space functions $G^0_i$ and $G^1_i$ are determined by the conditions
\bsubeq
\bea
&&G^0_i=C_\pfc\, ,                                             \lab{5.2a}\\
&&G^1_i+\{G^0_i, H_T\}=C_\pfc\, ,                              \lab{5.2b}\\
&&\{G^1_i,H_T\}=C_\pfc\,.                                      \lab{5.2c}
\eea
\esubeq
and $C_\pfc$ denotes a primary FC constraint. The construction starts with any primary FC constraint $G^0_i$, and the algorithm describes how the corresponding $G^1_i$ should be determined.

\subsection{Gauge generator of the generic TG}

The total and canonical Hamiltonians of TG are given by
\bsubeq\lab{5.3}
\bea
&&H_T=H_c+u^i{_0}\pi_i{^0}+\pd_\a D^\a\, ,                         \\
&&H_c:=N\cH_\orth+N^\a\cH_\a=\vth^i{_0}\cH_i\,,
\eea
where:
\be
\cH_i:=n_i\cH_\orth+e_\bi{^\a}\cH_\a\,.
\ee
\esubeq
The PB algebra of the Hamiltonian constraints \eq{4.1}, obtained by the Legendre transform technique \cite{PM19} can be transformed to an equivalent form as
\be
\{\cH_i,\cH_j\}=T^n{}_{ij}\cH_n\,\d\,.                          \lab{5.4}
\ee

Since the only primary FC constraints in TG are $\vphi_i=\pi_i{^0}$, we start the construction of the gauge generator by taking
\be
G^0_i=-\pi_i{^0}\,.
\ee
The condition \eq{5.2b}, combined with $\{ G^0_i,H_T\}=\cH_i$, implies
\be
G^1_i=-\cH_i+\a_i{^m}\pi_m{^0}\, ,
\ee
where the unknown coefficients $\a_i{^m}$ are determined by \eq{5.2c},
\bsubeq
\be
\{G^1_i,H_T\}=-\vth^k{_0}T^n{}_{ik}\cH_n\d-\a_i{^n}\cH_n\d=C_\pfc\, .
\ee
Solving this condition for $\a_i{^n}$ yields
\be
\a_i{^n}=-\vth^k{_0}T^n{}_{ik}\,,\qquad
              G_i^1=-\cH_i-\vth^k{_0}T^n{}_{ik}\pi_n{^0}\, ,
\ee
\esubeq
and the final gauge generator takes the form
\be
G=-\dot\xi^i\pi_i{^0}-\xi^i(\cH_i+\vth^k{_0}T^n{}_{ik}\pi_n{^0})\,.
\ee

It is convenient to introduce the coordinate components of $\xi^i$ by $\xi^i=\vth^i{_\m}\xi^\m$, so that
\bsubeq
\bea
&&G=-\dot\xi^\m \vth^i{_\m}\pi_i{^0}-\xi^\m P_\m\, ,              \\
&&P_\m=\vth^i{_\m}(\cH_i+\vth^k{_0}T^n{}_{ik}\pi_n{^0})+\pi_i{^0}\pd_\m \vth^i{_0}
    =\vth^i{_\m}\cH_i+\pi_i{^0}\pd_\m \vth^i{_0}\,.
\eea
\esubeq
Using the on shell relation $\pd_0 \vth^i{_0}=u^i{_0}$, one obtains
\bsubeq\lab{5.10}
\bea
&&P_0=H_c+u^i{_0}\pi_i{^0}=H_T-\pd_\a D^\a\, ,                    \\
&&P_\a=\cH_\a+\pi_i{^0}\pd_\a \vth^i{_0}
      =\pi_i{^\m}\pd_\a \vth^i{_\m}-\pd_\b(\pi_i{^\b}\vth^i{_\a})\,.
\eea
\esubeq
This form of $G$ correctly reproduces the local translations as a symmetry of the generic TG \cite{PM19}.

\subsection{Generalization to \mb{f(\mT)} gravity}

In the formalism  of Section \ref{sec5}, the dynamical Hamiltonian is defined by the Legendre transform, which means that $C_{ij}$ is included in $\cH_\orth$. Then, the total/canonical Hamiltonians can be written in the form
\eq{4.4}, representing an isomorphic image of the TG formulas \eq{5.3}. In particular, the structure functions of the PB algebra \eq{4.5} in $f(\mT)$ gravity are identical to those of the PB algebra \eq{4.1} in TG. Hence, the Castellani procedure is practically identical to the one used in TG. As a result, the gauge generator in $f(\mT)$ gravity is found to be
\be
\bG\approx
  -\dot\xi^i\pi_i{^0}-\xi^i(\bcH_i+\vth^k{_0}T^n{}_{ik}\pi_i{^0})\,.\lab{5.12}
\ee
Here, the weak equality is a consequence of the weak equalities appearing in \eq{4.5}. However, one can show that the weak equality can be safely replaced by the strong one, so that
\bsubeq\lab{5.13}
\bea
&&\bG=-\dot\xi^\m \vth^i{_\m}\pi_i{^0}-\xi^\m P_\m\, ,                 \\
&&P_0=\bH_c+u^i{_0}\pi_i{^0}=H_T-\pd_\a D^\a\, ,                       \\
&&P_\a=\bcH_\a+\pi_i{^0}\pd_\a \vth^i{_0}
      =\pi_i{^\m}\pd_\a \vth^i{_\m}-\pd_\b(\pi_i{^\b}\vth^i{_\a})+\pi_\phi\pd_\a\phi\,.
\eea
\esubeq
Indeed, by comparing this result with the one displayed in \eq{5.10}, one can conclude that the gauge generator $\bG$ produces the correct local translations when acting on the phase-space variables $(\vth^i{}_\m,\pi_i{}^\m)$. Moreover, a direct verification shows that its action on $(\phi,\pi_\phi)$ is also correct. Hence, $\bG$ acts correctly on the whole phase space of $f(\mT)$ gravity.

\bitem
\item The gauge generator $\bG$ is constructed by assuming that $\pi_i{}^0$ is FC and using the PB algebra \eq{4.5}. The fact that $\bG$ is the true gauge generator of $\fT$ gravity implies that the Hamiltonian constraints $(\bcH_\orth,\bcH_\a)$ must be FC, independently of the properties of other constraints, like $\pi_\phi,C_{ij}$ or $\chi_\phi,\chi_{ij}$.
\eitem
\section{Determining the multipliers (\mb{u_\phi,u^{ij}})}\label{sec6}
\setcounter{equation}{0}

Preservation of the primary constraints $\pi_\phi$ and $C_{mn}$ leads to the conditions \eq{3.16} and \eq{3.17}, respectively, which either produce new constraints or determine some multipliers (at least generically). These conditions can be written in the form
\bsubeq\lab{6.1}
\bea
\chi_\phi:&& u_{\orth\bj}F^{\orth\bj}+\frac{1}{2}u_{\bi\bj}F^{\bi\bj}
                                           \approx NF_\phi\,,   \lab{6.1a}\\
\chi_{ij}:&&F_{ij}\bu_\phi
    +2a_0J\phi_\bk\,\d^{\orth\bk\bn}_{ijm}u^m{_n}\approx X_{ij}\,,\lab{6.1b}
\eea
\esubeq
where
\bsubeq\lab{6.2}
\bea
&&\bu_\phi:=N\pd_\orth\phi=u_\phi-N^\b\pd_\b\phi\, ,\qquad
  \phi_\bk:=\pd_\bk\phi \,,                                     \lab{6.2a}\\
&&X_{ij}:=N\phi_\bk\Big[\phi^{-1}g^{\bk\bm}
           (n_i\hpi_{(\bm\bj)}-n_j\hpi_{(\bm\bi)})
           +a_0J\d_{\bi\bj\br}^{\bk\bm\bn}T^\br{}_{\bm\bn}\Big]\, .
\eea
\esubeq
Making a 1+3 decomposition of \eq{6.1b}, one finds
\bea
&&F_{\orth\bj}\bu_\phi-Z_\bk u^\bk{_\bj}=X_{\orth\bj}\,,           \nn\\
&&F_{\bi\bj}\bu_\phi
+Z_\bk\big(\d^\bk_\bi u^\orth{_\bj}-\d^\bk_\bj u^\orth{_\bi}\big)
                                             =X_{\bi\bj}\, ,      \lab{6.3}
\eea
where $Z_\bk:=2a_0J\phi_\bk$. Then, using the notation
\be
F_\bk:=\frac{1}{2}\ve_{\bk\bm\bn}F^{\bm\bn}\, ,\qquad
u_\bk:=\frac{1}{2}\ve_{\bk\bm\bn}u^{\bm\bn}\, ,                   \lab{6.4}
\ee
the system of seven equations \eq{6.1} for the seven unknown multipliers $(\bu_\phi,u_{\orth\bk},u_\bk)$ takes the form
\bsubeq\lab{6.5}
\bea
&&F^{\orth\bn}u_{\orth\bn}+F^\bn u_\bn\approx NF_\phi\, ,       \lab{6.5a}\\
&&F^{\orth\bm}\bu_\phi-Z_k\ve^{\bk\bm\bn}u_\bn=X^{\orth\bm}\,,  \lab{6.5b}\\
&&F^\bm\bu_\phi-Z_k\ve^{\bk\bm\bn}u_{\orth\bn}=X^\bm\,.         \lab{6.5c}
\eea
\esubeq
For an extension to $D$ spacetime dimensions, see Appendix \ref{appF}.

Further analysis is organized by separating two complementary cases, $\phi_\bk\ne 0$ and $\phi_\bk=0$.

\subsection{\mb{\phi_\bk\ne 0}}

The contraction of Eqs. \eq{6.5b} and \eq{6.5c} with $\phi_\bm$ yields
\bsubeq\lab{6.6}
\bea
&&\phi_\bm F^{\orth\bm}\bu_\phi=\phi_\bm X^{\orth\bm}\,,       \lab{6.6a}\\
&&\phi_\bm F^\bm\bu_\phi=\phi_\bm X^\bm\, .                    \lab{6.6b}
\eea
\esubeq
These conditions have two important consequences. First, they \emph{generically} determine $\bu_\phi$,
\bsubeq
\be
\Big[(\phi_\bm F^{\orth\bm})^2+(\phi_\bm F^\bm)^2\Big]\bu_\phi
   =(\phi_\bm F^{\orth\bm})\phi_\bn X^{\orth\bn}
   +(\phi_\bm F^\bm)\phi_\bm X^\bm\,,
\ee
as long as\footnote{Here one can explicitly see that, as argued qualitatively in \cite{Chen1998,Chen2014}, the $\fT$ theory is indeed vulnerable to problems with nonlinear constraints leading to multipliers which can become unbounded for certain field values. This is an indication of a tachyonic propagating mode:  in this case when \eq{6.7b} approaches zero $\bu_\phi$ becomes unbounded---unless the right-hand sides of \eq{6.6a} and \eq{6.6b} also vanish.}
\be
(\phi_\bm F^{\orth\bm})^2+(\phi_\bm F^\bm)^2\ne 0\,.            \lab{6.7b}
\ee
\esubeq
This is our main generic assumption, which implies that at least one of the two terms $\phi_\bm F^{\orth\bm}$ and $\phi_\bm F^\bm$ does not vanish. And second, they produce a \emph{new secondary constraint},
\be
\chi:=\ub{\phi_\bn F^{\orth\bn}(\phi_\bm X^\bm)}_{A_1}
      -\ub{\phi_\bn F^\bn(\phi_\bm X^{\orth\bm})}_{A_2}\,.       \lab{6.8}
\ee
A more detailed expression for $\chi$ is obtained using the identities
\bea
&&F_{\orth\bj}\equiv 2a_0J \bV_\bj\,,\qquad
  F_{\bi\bj}\equiv -2a_0J T^\orth{}_{\bi\bj}\, , \qquad
  F^\bk\equiv-a_0J\ve^{\bk\bm\bn}T_{\orth\bm\bn}\,,                  \nn\\
&&X_{\orth\bj}\equiv N\phi^{-1}\phi^\bm\hpi_{(\bm\bj)}\,,\qquad
  X_{\bi\bj}\equiv a_0 b\phi_\bk
       \d_{\bi\bj\br}^{\bk\bm\bn}T^\br{}_{\bm\bn}\,,\qquad
  X^\bi\equiv a_0b\phi_\bk\ve^{\bk\bm\bn}T^\bi{}_{\bm\bn}\,,          \nn
\eea
where $\bV_\bi:= T^\bk{}_{\bk\bi}$. As a consequence,
\bea
&&A_1=2a_0J\phi_\br\bV^\br\cdot
             a_0b\phi_\bi\phi_\bk\ve^{\bk\bm\bn}T^\bi{}_{\bm\bn}\,,   \nn\\
&&A_2=-a_0J\ve^{\bk\bm\bn}\phi_\bk T_{\orth\bm\bn}\cdot
       N\phi^{-1}\phi_\br\phi_\bi\pi^{(\br\bi)}\,,                    \nn\\
\Ra&&\chi=a_0b\ve^{\bk\bm\bn}\phi_\bk\phi_\bi\phi_\bj\phi^{-1}
     \big(T_{\orth\bm\bn}\hpi^{(\bi\bj)}
          +2a_0J\phi\bV^\bi T^\bj{}_{\bm\bn} \big)\,.             \lab{6.9}
\eea

To examine how this result affects the multipliers $u_{\orth\bk}$ and $u_\bk$, we split them into components parallel to and orthogonal to $\phi_\bk$,
\bea
&&u_{\orth\bk}=u_\orth\phi_\bk+\hu_{\orth\bk}\, ,\qquad
                                         \hu_{\orth\bk}\phi^\bk=0\,, \nn\\
&&u_\bk=u\phi_\bk+\hu_\bk\, ,\hspace{46pt} \hu_\bk\phi^\bk=0\, .
\eea
Returning to the general conditions \eq{6.5}, note first that Eqs. \eq{6.5b} and \eq{6.5c} contain only the orthogonal components $\hu_\bn$ and $\hu_{\orth\bn}$,
\bea
&&F^{\orth\bm}\bu_\phi-2a_0J\phi_\bk\ve^{\bk\bm\bn}\hu_\bn=X^{\orth\bm}\,,\nn\\
&&F^\bm\bu_\phi-2a_0J\phi_\bk\ve^{\bk\bm\bn}\hu_{\orth\bn}=X^\bm\, .
\eea
Then, substituting the solutions of these equations for $\hu_\bn$ and $\hu_{\orth\bn}$ into \eq{6.5a} yields one linear equation for the parallel components $u_\orth$ and $u$,
\bsubeq\lab{6.12}
\be
u_\orth (F^{\orth\bn}\phi_\bn)+u(F^\bn\phi_\bn)
              =NF_\phi-F^{\orth\bn}\hu_{\orth\bn}-F^\bn\hu_\bn\,.  \lab{6.12a}
\ee

The second equation for $u$ and $u_\orth$ is obtained from the term $\{C_{mn},\chi'\}$ in the preservation condition for $\chi$ (see Appendix \ref{appG}):
\bea
-\pd_0\chi'&=&\{H_T,\chi'\}
            =\frac{1}{2}u^{mn}\{C_{mn},\chi'\}+\text{multiplier independent terms}\nn\\
  &=&u_\orth\big(\phi^\bn\{C_{\orth\bn},\chi'\}\big)
   +u\Big(\frac{1}{2}\ve^{\bm\bn\bk}\phi_\bk\{C_{mn},\chi'\}\Big)+\text{multiplier independent terms}.\nn\\
\eea
\esubeq
The solutions of the system of linear equations \eq{6.12} for $u_\orth$ and $u$ depend on the form of the determinant
\be
D(x,x'):=F^{\orth\bn}\phi_\bn
  \Big(\frac{1}{2}\ve^{\bi\bj\bk}\phi_\bk\{C_{\bi\bj},\chi'\}\Big)    -(F^\bn\phi_\bn)\big(\phi^\bj\{C_{\orth\bj},\chi'\}\big)\, .     \lab{6.13}
\ee

To proceed further in a simple way, suppose that  $F^{\orth\bn}\phi_\bn\ne 0$, in accordance with our generic condition \eq{6.7b}. (The alternative case $F^\bn\phi_\bn\ne 0$ can be handled in a similar way.) Then, \eq{6.12a} can be interpreted as an equation that defines $u_\orth$ in terms of $u$. Next, introduce the notation
\bsubeq\lab{6.14}
\be
u_\orth\big(\phi^\bn\{C_{\orth\bn},\chi'\}\big)
   +u\Big(\frac{1}{2}\ve^{\bm\bn\bk}\phi_\bk\{C_{mn},\chi'\}\Big)
                                                       =:g(x')\,,  \lab{6.14a}
\ee
multiply this relation by $F^{\orth\bp}\phi_\bp$,
\be
 u_\orth(F^{\orth\bp}\phi_\bp)\big(\phi^\bn\{C_{\orth\bn},\chi'\}\big)
   + u(F^{\orth\bp}\phi_\bp)
   \Big(\frac{1}{2}\ve^{\bm\bn\bk}\phi_\bk\{C_{\bm\bn},\chi'\}\Big)= F^{\orth\bk}\phi_\bk g(x')\,,
\ee
\esubeq
insert the expression for $u_\perp$ determined by \eq{6.12a}, and rearrange to get
\be
u\left[F^{\orth\bp}\phi_\bp \Big(\frac{1}{2}\ve^{\bm\bn\bk}\phi_\bk\{C_{\bm\bn},\chi'\}\Big)
   -F^\bp\phi_\bp\big(\phi^\bn\{C_{\orth\bn},\chi'\}\big)\right]
   = \hbox{known terms}.
\ee
This equation for the last undetermined multiplier has the form
\be
u(x)D(x,x')=G(x')\,,                                               \lab{6.16}
\ee
where $D(x,x')$ is defined in \eq{6.13}.
In view of the derivatives of the $\d$ function buried in $D(x,x')$, see Appendix \ref{appG}, it is important to be mindful of the implicit integration over the variable $x$.  Carrying out the integrations by parts (and then in the end dropping the prime) will lead to a relation of the form\footnote{This is the first time that we have encountered \emph{a differential equation} for a multiplier. It seems strange to us.  How does this affect locality?}
\be
A^\g \pd_\g u+\a u=G\,.                                            \lab{6.17}
\ee

The explicit functional forms for $A^\g$ and $\a$ can be straightforwardly obtained from the explicit form of $\{C_{ij},\chi\}$, derived in Appendix \ref{appG}. Several scenarios are possible.

\bitem
\sitem{\emph{Generic scenario}. If the differential equation \eq{6.17} can be solved for $u$, all the multipliers are determined. Then the numbers of Lagrangian variables, first and second class constraints are, respectively, $N=16+1,~ N_1(\pi_i{^0},~\cH_i)=8,~ N_2(\pi_\phi,C_{ij},\chi)=8$, and consequently, the number of d.o.f. is $N^*=16+1-8-8/2=5$. In particular, such a scenario could be realized if $A^\gamma$ vanishes but $\a\ne0$. In that special case the relation \eq{6.17} degenerates to a \emph{linear algebra relation} for the final multiplier.}

\sitem{If $A^\gamma$ and $\alpha$ both vanish (this seems highly unlikely to us),  then $G$ is a new secondary which must be preserved.  That in turn could lead to further constraints with either the remaining multiplier being eventually determined or maybe remaining undetermined.  We do not see any easy way in principle to restrict the possible length of this constraint chain. If it is long enough there will be no d.o.f..}

\sitem{We cannot exclude some other, albeit unlikely, possibilities. Thus, for instance,  if $\pd_0\chi$ identically vanishes, there would be only one condition for the multipliers $u$ and $u_\orth$. Then, there would remain one undetermined multiplier, one degree of ``remnant local Lorentz symmetry" \cite{Chen2014}. As a consequence, one combination of the components of $(\pi_\phi,C_{ij})$ would be first class and would lead to the first class secondary $\chi$, and six components of $(\pi_\phi,C_{ij})$ would be second class. Hence, $N_1=8+2=10$, $N_2=6$, and $N^*=17-10-6/2=4$.}
\eitem
It seems likely that there are more than the three d.o.f. claimed by Ferraro and Guzm\'an \cite{Ferraro:2018tpu}, and no---or at most two---``remnant" local Lorentz symmetries, not five.

\subsection{Sector \mb{\phi_\bk=0}}

The phase space for field theories  has in general various sectors with distinct dynamics. One way this can happen is if one considers just the subset of initial data with some special symmetry (e.g., spherical, axisymmetry, homogeneous).  Another way is by restricting to the subset of fields where some quantities vanish.

The objective of the following discussion is to take a broader view on the phase-space constraint/multiplier story in the $\fT$ theory. We focus on the $\phi$-Lorentz sector with the primary constraints $\pi_\phi$ and $C_{ij}$. Preserving these constraints leads to the conditions \eq{6.5}. Clearly, there is a special sector ${\bar\G}$ of the whole phase space $\G$, defined by\footnote{There are different types of vanishing.  One case is \emph{instantaneous}, that is a quantity vanishes at $t=t_0$, but not at earlier or later instants.  Such a case need not be pursued, as one can just adjust the initial time a little to avoid the vanishing. (However, one should then be concerned that the evolution stays regular as one approaches the critical time \cite{Chen1998}.) One should instead focus on the cases where the system evolution stays on the subset where some quantity vanishes.  Another complication that could happen but cannot be treated generally is where a quantity vanishes on a subset of the spatial hypersurface.}
\be
\phi_{\bk}=0.                                                     \lab{6.18}
\ee
\emph{This is a very important sector}; it includes the homogeneous cosmologies where $\phi=\phi(t)$.\footnote{Note: interest in the $\fT$ theory is mainly as a potential solution to the dark matter/dark energy cosmological puzzles.}

The content of \eq{6.18} is clarified by the following observations.

The differential conditions $\pd_\bk\phi=0$ that define $\bar\G$ do not eliminate $\phi$ as a degree of freedom, they only restrict the coordinate dependence of $\phi$ (invariance under spatial translations). Hence, they do not change the dimension of the phase space.

Additional information on the restriction \eq{6.18} comes from its dynamical preservation,
\bsubeq\lab{6.19}
\be
\pd_0\phi_\bk:=\{\phi_\bk,H_T\}
  =\{\phi_\bk,u'_\phi\pi'_\phi+C'_{ij}(u^{ij})'/2\}
  =u'_\phi\pd_\bk\d-u'_\bk{}^\bj\phi_\bj\approx 0\,,
\ee
where we used the relations \eq{H.1}$_6$. By partial integration, one finds
\be
\pd_\bk u_\phi=0\,,
\ee
\esubeq
which is just  a consistent extension of the condition \eq{6.18} on $\phi$ to an analogous condition on $u_\phi=\pd_0\phi$.

In the sector $\bar\G$, the relation \eq{6.2a} simplifies into
\be
\bu_\phi=u_\phi\,,                                                 \lab{6.20}
\ee
and the relations \eq{6.5} reduce to
\bsubeq\lab{6.21}
\bea
F^{\orth\bn}u_{\orth\bn}+F^\bn u_\bn&\approx& NF_\phi\, ,             \\
F^{\orth\bm}u_\phi&=&0\,,                                             \\
F^\bm u_\phi&=&0\,.
\eea
\esubeq
In contrast to \eq{6.5}, the conditions \eq{6.21} do not produce any additional constraint $\chi$. They allow one to find solutions for the canonical parameters by making some specific assumptions on the coefficients $F^{\orth\bm},F^\bm$. Our first case is described by the following scenario.
\bitem
\sitem{If at least one of $F^{\orth\bm},F^\bm$ is nonvanishing, then $u_\phi=0$ and, as a consequence, $\pd_0\phi=\{\phi,H_T\}=\{\phi,u'_\phi\pi'_\phi\}=0$. Combining this result with $\phi_\bk=0$, one finds that $\phi$ must be a true constant, $\phi=c$. As a consequence, the dynamical content of the Lagrangian field equation \eq{2.8} takes the \tgr\ form, up to a cosmological constant term $V(c)/c$.}
\eitem
According to Eq. \eq{2.7b}, $\phi=c$ implies that $\mT$ is also a constant. Thus, any constant $\mT$ configuration allows the existence of nontrivial solutions provided they are also solutions of \tgr, which is a rather strong, nonperturbative restriction on the $\fT$ dynamics in the ``cosmological" sector $\phi_\bk=0$.

To complete our canonical analysis, let us now consider the number of d.o.f..
If the premise of (s4) holds, the theory reduces to \tgr\ with 2 d.o.f..
The content of (s4) implies that the conditions \eq{6.21} have just one other physically relevant realization:
\bsubeq\lab{6.22}
\bea
&&F_{\orth\bk}\equiv 2a_0J \bV_\bk =0\, ,                        \lab{6.22a}\\
&&F^\bk\equiv -a_0J\ve^{\bk\bm\bn}T_{\orth\bm\bn}=0\, ,          \lab{6.22b}\\
\Ra&&F_\phi\equiv \frac{1}{2a_0\phi^2}P^2+J(\mT-\pd_\phi V)=0\,. \lab{6.22c}
\eea
\esubeq
Since these conditions restrict the dimension of the phase space, one should impose their dynamical preservation.

Taking into account the relation $\{J,\cH'_{mn}\}=0$, see \eq{H.1}$_3$, the preservation of $F_{\orth\bn}$ becomes equivalent to (integration over $x'$ implicit)
\bea
\pd_0 \bV_\bn=\{\bV_\bn,H_T\}=
    \{\bV_\bn,\cH'_{ij}\}\frac{1}{2}(u^{ij})'+\{\bV_\bn,\wH_c\}\approx 0\,,
\eea
where we used the expression \eq{3.10} for $H_T$. A direct inspection of the second term shows that it does not depend on the canonical multipliers.  When the above relation, with interchanged $x$ and $x'$, is combined with \eq{H.3}$_4$, it yields
\bea
\pd_0\bV'_\bn&=&-\frac{1}{2}u^{ij}\{\cH_{ij},\bV'_n\}+\b'_\bn        \nn\\
   &\approx& -\big(u^{\bi\bj}T_{\bj\bi\bn}\d +u^\bi{}_n\bV_\bi
                   -u^{\orth\bj}T_{\orth\bj\bn}\big)\d
-(e_\bi{^\b}e_{\bn}{^\g})'\big[\pd'_\b(u^i{_\g}\d)
                                 -\pd'_\g(u^i{_\b}\d\big]+\b'_\bn\,,  \nn
\eea
where $\b'_n:=-\{\wH_c,\bV'_\bn\}$. Integrating over $x$, and replacing $x'$ by $x$ in the final result, one obtains three differential conditions on the six multipliers $u^{ij}$,
\bsubeq\lab{6.24}
\be
-\big(u^{\bi\bj}T_{\bj\bi\bn}+u^\bi{}_n\bV_\bi-u^{\orth\bj}T_{\orth\bj\bn}\big)
    +(e_\bi{^\b}e_{\bn}{^\g})(\pd_\b u^i{_\g}
               -\pd_\g u^i{_\b})+\b_\bn\approx 0\,.               \lab{6.24a}
\ee

Similarly, the preservation of $F^\bn$ is equivalent to
\be
\pd_0 T_{\orth\bm\bn}\approx\{T_{\orth\bm\bn},\cH'_{ij}\}\frac{1}{2}(u^{ij})'
                           +\b_{\orth\bm\bn}\approx 0\,.            \nn
\ee
Then, relation \eq{H.3}$_2$ implies
\be
\pd_0 T'_{\orth\bm\bn}=-\Big[(u^{\orth\bj}T_{\bj\bm\bn}
   +u^\bi{}_mT_{\orth\bi\bn}-u^\bi{}_nT_{\orth\bi\bm})\d
   -n'_i(e_\bm{^\b}e_\bn{^\g})'\big[\pd'_\b(u^i{_\g}\d)
                     -\pd'_\g(u^i{_\b}\d)\big]\Big]+\b'_{\orth\bm\bn} \nn
\ee
and consequently,
\be
-u^{\orth\bj}\,T_{\bj\bm\bn}-u^\bi{}_mT_{\orth\bi\bn}+u^\bi{}_nT_{\orth\bi\bm}
+n_i(e_\bm{^\b}e_\bn{^\g})\big(\pd_\b u^i{_\g}
          -\pd_\g u^i{_\b}\big)+\b_{\orth\bm\bn}\approx 0\,.     \lab{6.24b}
\ee
Here, we have a set of three conditions on the six multipliers $(u^{\bi\bj},u^{\orth\bj})$.

Finally, the preservation of $F_\phi$ takes the form
\be
\pd_0 F_\phi=\{F_\phi,H_T\}=\{F_\phi, u'_\phi\pi'_\phi\}+\text{more}
            =u_\phi(\pd_\phi F_\phi)+\text{more}\approx 0\,.     \lab{6.24c}
\ee
\esubeq

Relations \eq{6.24a} and \eq{6.24b} are differential equations for the canonical multipliers $u^{ij}$, whereas the condition \eq{6.24c} determines $u_\phi$, provided $\pd_\phi F_\phi\ne 0$.

\bitem
\sitem{In the \emph{generic scenario}, relations \eq{6.24} determine the multipliers $(u_\phi,u^{ij})$. Then, in the phase space $\bar\G$ with 17 Lagrangian variables  $(\vth^i{}_\m,\phi)$, we have\\
  \ph{xxx} 3+3+1=7 new constraints \eq{6.22},\\
  \ph{xxx} seven preservation conditions \eq{6.24}, and seven determined multipliers $(u^{ij},u_\phi)$.\\
  Since the seven primary constraints $(\pi_\phi,C_{ij})$ and the seven new constraints \eq{6.22} are second class, and $N_1=8$, the number of physical d.o.f. is $N^*=17-8-7=2$, the same as in \tgr.}
\eitem
The case considered in (s5) is exactly the situation when the premise of (s4) does not hold. So (s4) leading to \tgr\ with 2 d.o.f. is not relevant.
The scenario (s5) also has 2 d.o.f., but we don't yet know its
dynamical content in detail. However, since the 8 first class
constraints $(\pi_i{}^0,\cH_i)$ are associated to diffeomorphisms, and the extra Lorentz constraints $C_{ij}$ are second class, it is clear that the case (s5) cannot be equivalent to \tgr.

\section{Summary and discussion}\label{sec7}
\setcounter{equation}{0}

In the present paper, we performed a detailed Hamiltonian analysis of $\fT$ gravity, with a focus on the local Lorentz invariance, the number of the physical d.o.f., and  the issue of nonlinear constraint effects.
Our main results can be summarized as follows.

(r1) The central role of the Lorentz constraint $C_{ij}$ with respect to the status of local Lorentz invariance can be seen already at an early stage of the canonical analysis. Namely, by showing that $\{C_{ij},C_{kl}\}$ does not vanish weakly, which means that $C_{ij}$ is not first class, one can directly conclude that local Lorentz invariance is broken.

(r2) To determine the classification of all the constraints and calculate the number of physical d.o.f., we found it convenient to first prove the first-class nature of the ADM components of the canonical Hamiltonian. This significantly simplifies further analysis and, as an ``aside" but expected result, it implies the diffeomorphism invariance of $\fT$ gravity.

(r3) The classification of the remaining constraints is based on the preservation conditions of the primary constraints $(\pi_\phi,C_{ij})$,
interpreted as seven conditions on the seven multipliers $(u_\phi,u^{ij})$.
Then, in a somewhat parallel processing procedure, we analyzed which of these multipliers are determined (that is, associated to second-class constraints) and what happens with secondary constraints, if they exist. Following such an approach, we found that generically, the number of physical d.o.f. is $N^*=5$. Note also that in the special case $\pd_\bk\phi=0$, $\fT$ gravity reduces to GR with a cosmological constant, with $N^*=2$.

(r4) We confirmed that $\fT$ gravity is indeed vulnerable to the effects associated with nonlinear second class constraints \cite{Chen1998,Chen2014}.  When the dynamical variables evolve toward values such that certain quantities approach zero, certain canonical multipliers can diverge---signaling an associated anomalous propagation.  Such behavior can be an indication of a fatal problem.

To gain a deeper insight into our results, we compare them to those of Li et al. \cite{LMM11} and Ferraro and Guzm\'an \cite{Ferraro:2018tpu}.

\prg{1.} The basic results of Li et al.  in $D=4$ are presented in Section 4 of Ref. \cite{LMM11}. Their Eqs. (25)--(28), representing the PB algebra involving the set of the $\phi$-Lorentz primary constraints $(\pi_\phi,C_{ij})$ and the canonical Hamiltonian $H_0$, are in complete agreement with our findings. In particular, their PB (25), with $G^{ab}$ given in the first line of the next page, is identical to our result for $\{C_{ij},C_{mn}\}$ in \eq{3.18}. On the other hand, it should be contrasted with the Lorentz PB algebra closure found in Eq. (70)$_1$ of Ref. \cite{Ferraro:2018tpu}. We did not find any comment by Ferraro and Guzm\'an on this  disagreement, although it is of essential importance for the Lorentz invariance and the counting of d.o.f..

Next, Li et al. continue with the analysis of the three equations (29) by interpreting them  as $1+6+1=8$ conditions for the seven multipliers.
In addition to the last two equations that we considered (seven conditions for the seven multipliers), they included here also the preservation condition for the canonical Hamiltonian $H_0$, equal to our $\wH_c$. In our approach, we gave a completely separate discussion of the preservation of $\wH_c$. Namely, in sections 4 and 5, we showed that a suitably modified canonical Hamiltonian   $\bH_c=\wH+(1/2)u^{mn}C_{mn}+u_\phi\pi_\phi$ is first class, which implies its preservation. Hence, Eq. (29)$_1$ is not really needed, it is just a consequence of the last two equations.

Without knowing that, the authors continue by writing the eight conditions (29) in the form of a homogeneous matrix equation with an $8\times 8$ antisymmetric matrix $M$ having a vanishing determinant and rank 6. The condition $\det M=0$ is written in the form of a new constraint $\pi_1:=\sqrt{\det M}\approx 0$, whose preservation  $\pd_0\pi_1\approx 0$ yields a new condition on the multipliers. Alternatively, by disregarding the redundant Eq. (29)$_1$, one is left with $6+1=7$ conditions for seven multipliers. As we showed in subsection 6.1, the second equation gives (generically) five conditions on multipliers plus a secondary constraint $\chi$, the preservation of which produces one more condition on the multipliers, $\pd_0\chi\approx 0$. Thus, the seven equations (29)$_2$ and (29)$_3$ could be written as a homogeneous matrix equation, with a $7\times 7$ matrix of rank 6. This confirms that Eq. (29)$_1$ is indeed superfluous. Using it does not do any serious harm, but it does complicate the analysis. Although both approaches in the \emph{generic} scenario predict the same number of d.o.f., $N^*=5$, our formalism is more explicit and practical.

At the end of section 5, Li et al. discuss the d.o.f. for a $D$-dimensional spacetime. The result of our analysis in Appendix \ref{appF}, $(D-1)$ d.o.f., agrees with their finding in the Lorentz $\phi$ sector. As a final remark, in discussing the results obtained from the second equation in $D$ dimensions, the authors write: ``One can check that in four dimensions the constraint derived from the second equation of eq. (29) and square root of the determinant of $M$ eq. (36) are exactly the same." This means that our $\chi$ coincides with their $\pi_1$. If we trust this assertion, then our secondary constraint result is ``exactly the same" as theirs.

We also note that, in their appendix, Li et al. find, just as we did, a first order differential equation for the last canonical multiplier.

\prg{2.} As we mentioned above, one of the main errors in Ferraro and Guzm\'an \cite{Ferraro:2018tpu} is their claim that the PB algebra of the constraints $G^{(1)}_{ab}$ in Eq. $(70)_1$ closes just like the ordinary Lorentz algebra, which is in contradiction to our result \eq{3.18} [and Li et al. Eq. (25)]. This error seriously affects their analysis, leading them to claim that five of the six constraints $G^{(1)}_{ab}$ are first class, not second class.

There are, however, a number of other errors, but we choose to comment here on only two of them. We begin by noting that the last equality in their (65) implies $F_\phi=0$. Indeed, as shown in our Appendix \ref{appB}, the Hamiltonian transcript of $E(\mT-\pd_\phi V)$ weakly vanishes.
Then, since $F_\phi$ introduced in their Eq. (64) does not vanish, the last equality in (65) cannot be correct.

Moreover, Ferraro and Guzm\'an calculated the preservation condition for $G^{(1)}_{ab}$ in their Eq. (81)$_2$. By comparison to our Appendix \ref{appC}, their result is recognized just as a fraction of the complete result, associated to our coefficient $B_4$.

Ferraro and Guzm\'an have published several follow-up works \cite{Ferraro:2018axk}, which have already attracted considerable attention. As they were based on the unsound foundation \cite{Ferraro:2018tpu},  they are not reliable guides.

\bigskip
In this paper we presented a detailed analysis of the puzzling $\fT$ Hamiltonian/constraint/d.o.f. issues.  This could be used as a solid foundation for certain future investigations into the nature of this curious theory.

\appendix
\section{Dynamical interpretation of the multipliers}\label{appA}
\setcounter{equation}{0}

Using the relations
\bsubeq\lab{A.1}
\bea
&&\{ \vth^k{_\g},\wcH'_\orth\}=\frac{1}{2a_0J\phi}\Big(
    \hpi^{(\bk\bn)}-\frac{1}{2}g^{\bk\bn}\hpi\Big)\vth_{n\g}\d
    -(n^k\pd_\g)'\d\,,                                            \lab{A.1a}\\
&&\{ \vth^k{_\g},\cH_\b\}=T^k{}_{\b\g}\d-(\vth^k{_\b}\pd_\g)'\d\,,       \lab{A.1b}\\
&&\{ \vth^k{_\g},C_{ij}\}= 2\d^k_{[i}\vth_{j]\g}\d\, ,
\eea
\esubeq
where $\hpi:=\hpi^\bk{}_\bk$,
one finds that the dynamical equation for $\vth^k{_\g}$ takes the form
\be
\pd_0 \vth^k{_\g}=\{ \vth^k{_\g},H_T\}=N\frac{1}{2a_0J\phi}\Big(
    \hpi^{(\bk\bn)}-\frac{1}{2}g^{\bk\bn}\hpi\Big)\vth_{n\g}
    +N^\b T^k{}_{\b\g}+\pd_\g \vth^k{_0}+u^{kn}\vth_{n\g}\, ,
\ee
where the integration over $d^3 x'$ is understood, and we used
$Nn^k+N^\b \vth^k{_\b}=\vth^k{_0}$. Based on the identity $\pd_0 \vth^k{_\g}-\pd_\g \vth^k{_0}\equiv NT^k{}_{\orth\g}+N^\b T^k{}_{\b\g}$, this equation leads to the interpretation of the multipliers $u^{mn}$ as displayed in \eq{3.11b}. Moreover, it implies
\be
\hpi^{(\bm\bn)}-\frac{1}{2}g^{\bm\bn}\hpi=2a_0J\phi T^{(m\orth\bn)}
\quad\Lra\quad
    \ir{T}\hpi^{\bm\bn}=2a_0J\phi\,\ir{T}T^{\bm\orth\bn}\,.   \lab{A.3}
\ee

\section{Lagrangian expression for \mb{\chi_\phi}}\label{appB}
\setcounter{equation}{0}

By rewriting the identity \eq{3.8a} in the form
\bsubeq
\be
J\cL-J\bcL=-\frac{1}{2a_0\phi}P^2+\hpi^{i\bm}T_{i\orth m}\,. \lab{B.1a}
\ee
and adding $J\phi(\bT-\pd_\phi V)$ to both sides, one obtains
\be
J\phi(\mT-\pd_\phi V)=-\frac{1}{2a_0\phi}P^2+J\phi(\bT-\pd_\phi V)
                   +\hpi^{i\bm}T_{i\orth\bm}\, .             \lab{B.1b}
\ee
\esubeq
Then, transforming the last term as
\be
\hpi^{i\bm}T_{i\orth m}\approx\frac{1}{a_0\phi}P^2
                          +\frac{1}{N}\hpi^{i\bn}u_{i\bn}\,,    \lab{B.2}
\ee
and using $\cH_{mn}\approx -\phi F_{mn}$, one obtains
\be
\vth\phi(\mT-\pd_\phi V)\approx \phi NF_\phi
                        +\frac{1}{2}u^{mn}\cH_{mn}\approx\phi\,\chi_\phi\,.
\ee

As a by-product of the above analysis, one can combine the identity \eq{B.1a} with the relation \eq{B.2} to obtain
\be
\vth\cL\approx\frac{N}{2a_0\phi}P^2+\vth\bcL
                     +\frac{1}{2}\ir{A}T^{m\orth n}\cH_{mn}\,.  \lab{B.4}
\ee

\section{The preservation of \mb{C_{ij}}}\label{appC}
\setcounter{equation}{0}

To calculate the preservation condition $\pd_0C_{ij}\approx 0$, we start from the relation
\be
\chi_{ij}:=\{C_{ij},H_T\}=\ub{\{C_{ij},N\cH_\orth\}}_{B_1}
  +\ub{\{C_{ij},N^\b\cH_\b\}}_{B_2}
  +\ub{\{C_{ij},(1/2)u^{mn}C_{mn}\}}_{B_3}
  +\ub{\{C_{ij},u_\phi\pi_\phi\}}_{B_4}\, ,                       \lab{C.1}
\ee
where we used $\{C_{ij},\pi_i{^0}\}\approx 0$. The $B_n$ terms are given by
\bea
B_1&=&-\frac{1}{2}\pd_\a\big[N(n_iC_{jk}-n_jC_{ik})e^{\bk\a}\d\big] \nn\\[2pt]
  &&-\frac{\pd_\a\phi}{\phi}N(n_i\hpi_{(j\bk)}-n_j\hpi_{(i\bk)})e^{\bk\a}\d
  -a_0(\pd_\a\phi)\ve_{ijmk}^{0\a\b\g}T^m{}_{\b\g}Nn^k\d\,,         \nn\\[2pt]
B_2&=&\pd_\b(N^\b C_{ij}\d)-(N^\b\pd_\b\phi)a_0B_{ij}\d\,,          \nn\\
B_3&=&u^{mn}(g_{jm}C_{in}-g_{im}C_{jn})\d
 +a_0u^{mn}\pd_\a\phi\big(g_{j\bn}B_{im}^{0\a}
      -g_{i\bn}B_{jm}^{0\a}\big)\d\,,                               \nn\\
B_4&=&a_0B_{ij}u_\phi\d=a_0B_{ij}(N\pd_\orth\phi+N^\b\pd_\b\phi)\,.
\eea
Then, transition to the weak equality yields
\bea
&&B_1\approx -\frac{\pd_\a\phi}{\phi}N(n_i\hpi_{(j\bk)}-n_j\hpi_{(i\bk)})e^{\bk\a}\d
  -a_0(\pd_\a\phi)\ve_{ijmk}^{0\a\b\g}T^m{}_{\b\g}Nn^k\d\,          \nn\\
&&B_3\approx a_0u^{mn}\pd_\a\phi\big(g_{j\bn}B_{im}^{0\a}
      -g_{i\bn}B_{jm}^{0\a}\big)\d\, ,                              \nn\\
&&B_4+B_2\approx a_0B_{ij}N\pd_\orth\phi\,.
\eea
After transforming the second term in $B_1$ and the whole of $B_3$ with the help of the identities
\bea
&&\vth^k{_\a}\ve_{ijmr}^{\a 0\b\g}T^m{}_{\b\g}Nn^r
  =-2b(T^\bk{}_{\bi\bj}-\d^\bk_\bi\bV_\bj+\d^\bk_\bi\bV_\bi)
  =-b\d^{\bk\bm\bn}_{\br\bi\bj}T^\br{}_{\bm\bn}\, ,                \nn\\
&&\vth^k{_\a}u^{mn}\big(g_{j\bn}B_{im}^{0\a}-g_{i\bn}B_{jm}^{0\a}\big)
          =-2Ju^{mn}(g_{\bj n}\d^{\orth\bk}_{im}
                       -g_{\bi n}\d^{\orth\bk}_{jm})
          =-2J\d^{\orth kn}_{imj}u^m{_n}\,,\qquad
\eea
the expression for $\chi_{ij}$ can be transformed exactly to the form \eq{3.17}.

\section{Direct Hamiltonian analysis for \mb{f(\mT)}}\label{appD}
\setcounter{equation}{0}

In this appendix, we examine what one gets if one tries to directly construct the $f(\mT)$ theory Hamiltonian. The $f(\mT)$ theory has the Lagrangian
\be
\tilde{\cal L}_{fT}=\vth{\cal L}_{fT}=\vth f(\mT),
\ee
where $\mT$ is the \tgr\ expression, displayed in (2.4).

The conjugate momenta come from
\begin{equation}
\pi_i{}^\mu:=\frac{\partial\big[\vth f(\mT)\big]}{\partial \dot \vth^i{}_\mu}
   =f'(\mT)\vth\cH_i{}^{0\mu}\,.
\end{equation}
Once again, one finds the \emph{sure} primary constraints
\be
\pi_i{}^0\approx0.
\ee
We find it convenient to represent the parallel momenta in a suggestive form
\be
\hpi_{i\bk}=\Phi J\cH_{i\orth\bk}\,,\qquad \Phi:=f'(\mT)\,.
\ee
The momenta $\pi_i{}^{\bar k}$ can again be split into irreducible components to give
\bea
p_{\orth\bk}:=\hat\pi_{\bot \bar k}/J&=&
                           -2\Phi T^{\bar m}{}_{\bar m\bar k}\,,   \\
\ir{A}p_{\bi\bk}:=\ir{A}\hat\pi_{\bar i\bar k}/J&=&
                             \Phi T_{\bot \bar i\bar k}\,,     \nn\\[5pt]
p^\bm{}_\bm:=\hat\pi^{\bar m}{}_{\bar m}/J&=&
                      -4\Phi T^{\bar m}{}_{\bot \bar m}\,,\label{Ptr} \\
\ir{T}p_{\bi\bk}:=\ir{T}\hat\pi_{\bar i\bar k}/J&=&
                              2\Phi\ir{T}T_{\bar i\bot\bar k}\,.\nn
\eea
Let us now  combine the first two relations in the form
\bsubeq
\bea
&&p_{ik}:=\frac{1}{2J}(\pi_{i\bar k}-\pi_{k \bar i})
      =\Phi\hF_{ik}\,,                                    \label{pF}  \\
&&\hF_{ik}:=(T^\orth{}_{\bi\bk}-n_i T^\bm{}_{\bm\bk}+n_k T^\bm{}_{\bm\bi})\,,
\eea
\esubeq
where both $p_{ik}$ and $\hF_{ik}$ are antisymmetric objects.
From this, it follows that
\be
\hF\cdot p:=\hF^{ik}p_{ik}=\Phi\hF^2, \quad
          \hF^2:=\hF\cdot\hF:=\hF^{ik}\hF_{ik}.       \label{FpfprimeTF2}
\ee
Clearly, vanishing $\hF^2$ is a special case.  Let us put this case aside for separate investigation, and consider the \emph{generic} case where $\hF^2$ does not vanish \emph{anywhere}. (One could also make a more complicated, ``less covariant'' analysis by considering the vanishing of $\hF_{\bot \bar k}$ and $\hF_{\bar i \bar k}$ separately.)
Then, we find from (\ref{FpfprimeTF2}) the component of (\ref{pF}) along $\hF_{ik}$,
\be
\Phi\equiv f'(\mT)=\frac{\hF\cdot p}{\hF^2}\,.              \label{fprimeT}
\ee

Using (\ref{fprimeT}), one can invert the relations (\ref{Ptr}) for some of the ``velocities'':
\be
T^{\bar m}{}_{\bot\bar m}=-\frac{p^{\bar m}{}_{\bar m}}{4\Phi}\,,\qquad
\ir{T}T_{\bar i\bot\bar k}=\frac{\ir{T}p_{\bar i\bar k}}{2\Phi}\,.
\ee
Furthermore, assuming $f''(\mT)\ne0$, by the implicit function theorem, the relation (\ref{fprimeT}) can be inverted to give
\be
\mT=(f')^{-1}(\Phi)\,.
\ee
With this relation, one can find the ``missing'' ``antisymmetric velocity''--momentum relation for the one velocity component along $F_{ik}$ from
\bea
\mT&\simeq&{\textstyle\frac12}\hbox{velocity}^2
                        +(\hbox{anti-sym velocity comp})^1+\bar\mT  \nn\\
&=&\frac{1}{2\Phi^2}P^2+\hF^{ik}\ir{A}T_{i\bot k}+\bar\mT\,,\label{Tfprime}
\eea
where $\bar\mT$ and $P^2$ are defined as in (3.9).  No other components of the ``antisymmetric velocity'' can be inverted for momenta; instead, the other five components of (\ref{pF}) are \emph{new primary constraints}, which can be written as\footnote{
An alternative form is
\be
\check{C}_{ik}:=\hF^2p_{ik}-(\hF\cdot p)\hF_{ik}\approx 0\, . \nn
\ee}
\be
\hat{C}_{ik}:=p_{ik}-\Phi\hF_{ik}\approx0\,,
                       \qquad \hat{C}_{ik}\hF^{ik}=0\,.       \label{npc}
\ee

It is interesting to point out certain structural similarities of the above results to those obtained in the $\phi\mT$ formalism of section \ref{sec3}: first, the expression (\ref{npc}) is an analogue of the extra primary constraint \eq{3.5}, and second, the relation (\ref{Tfprime}) is a counterpart of \eq{B.4}.

Now, one can construct the Hamiltonian. The \emph{total} Hamiltonian has the form
\be
{\cal H}_T={\cal H}_c+u^i{}_0\pi_i{}^0
   +{\textstyle\frac12}\hat u^{ik}\hat{C}_{ik},             \label{fTham}
\ee
including 4+5 \emph{primary} constraints with the canonical multipliers.
The explicit form of ${\cal H}_c$ can be found by following a close analogy to the procedure described in the main text, but the alternative $\phi\mT$ formalism seems to be much more practical.  Nevertheless, we want to stress that one could develop a complete Hamiltonian analysis based on \eqref{fTham}. Unlike the case of the six Lorentz generators of GR$_{\|}$, one will now find that the Poisson brackets algebra among the five primary constraints $\hat C_{ik}$ does not close, which is related to the fact that the Lorentz Lie algebra does not have a five-dimensional Lie subalgebra. From the analysis of the $\phi\mT$ formulation discussed in detail in the text, we can infer that the five constraints \eqref{npc} will be second class. If one goes further, one will find that the preservation of these five constraints will generically lead to four conditions on the five multipliers ${\hat u}^{ik}$ plus one secondary constraint $\chi$.  The preservation of the latter will, generically, yield a first order differential equation for the last multiplier. Generically, the number of physical d.o.f. is then
\be
N^*=N-N_1-N_2/2=16-8-6/2=5=2+3.
\ee

\section{Derivation of the algebra \eq{4.5}}\label{appE}
\setcounter{equation}{0}

Let us start from the observation that the PB algebra of the Hamiltonians $(\cH_\a,\cH_\orth)$ in \tgr\ has the form \eq{4.1}, as shown in \cite{PM19}. Then, since the result was derived with the Legendre transform representation for $\cH_\orth$ which does not depend on the explicit form of the Lagrangian, one can conclude that \eq{4.1} holds also in $f(\mT)$ gravity. Indeed, the presence of the variable $\phi$ in the Lagrangian has no effect on this algebra since none of the Hamiltonians $(\cH_\a,\cH_\orth)$ depends on $\pi_\phi$.
Knowing that, we will now show that the PB algebra \eq{4.5} for the $\phi$-modified Hamiltonians \eq{4.2} follows from \eq{4.1}. The proof is presented in three steps.

a1) Since $\cH_\a$ does not depend on $\phi$, the relation \eq{4.5a} follows directly from \eq{4.1a} and the definition of $\bcH_\a$.

a2) The dynamical Hamiltonian $\cH_\orth$ does not depend on $\pi_\phi$, whereas its $\phi$ dependence, combined with relations \eq{3.15} and \eq{3.12}, implies
\be
\{\pi_\phi,\cH_\orth\}=-\pd_\phi\cH_\orth=\chi_\phi\d\,.          \lab{E.1}
\ee
As a consequence,
\be
\{\bcH_\a,\bcH'_\orth\}\approx\{\cH_\a,\bcH'_\orth\}
                      \approx\{\cH_\a,\cH'_\orth\}\,,
\ee
which proves \eq{4.5b}.

a3) Relying on Eq. \eq{E.1}, one finds
\bsubeq
\bea
\{\bcH_\orth,\bcH'_\orth\}&=&\{\cH_\orth,\cH'_\orth\}
  +\{\cH_\orth,\pi'_\phi\pd'_\orth\phi'\}
      +\{\pi_\phi\pd_\orth\phi,\cH'_\orth\}                         \nn\\
  &&+\{\pi_\phi\pd_\orth\phi,\pi'_\phi\pd'_\orth\phi'\}
                \approx\{\cH_\orth,\cH'_\orth\}\,,
\eea
\esubeq
which confirms \eq{4.5c}.

\section{Solving for the multipliers in dimension \mb{D}}\label{appF}
\setcounter{equation}{0}

This is an alternative analysis to the one in Section \ref{sec6}, it is hardly more complicated and  extends the result to $D$ spacetime dimensions.

The equations to be solved are displayed in \eq{6.1a} and \eq{6.3}:
\bea
F^{\orth \bj}u_{\orth \bj}-\frac12F_{\bi\bj}u^{\bi\bj}+NF^\phi &\approx& 0, \label{one}\\
F^{\perp\bj}\bu_\phi- Z_{\bk}\d^{\bk\bj}_{\bl\bm}\frac12u^{\bl\bm}
  &\approx& X^{\perp\bj}\,,                                 \label{two}\\
F_{\bi\bj}\bu_\phi+ Z_{\bk}\d^{\bk\bl}_{\bi\bj}u_{\orth\bl}
  &\approx& X_{\bi\bj}\,.                                   \label{three}
\eea
These equations have, respectively, 1, $D-1$, $(D-1)(D-2)/2$ components.

The first relation gives \emph{one} restriction on $u^{ij}$, let us set it aside for now.
The component of the second relation projected along $\phi_{\bj}$ is
\be
(\phi_{\bj}F^{\perp\bj})\bu_\phi\approx\phi_{\bj}X^{\perp\bj}\,.\label{uphi}
\ee
\emph{Generically} (i.e., when $\phi_{\bj}F^{\perp\bj}\ne0$)\footnote{
 In the $D=4$ case in the main text, we considered the special cases $\phi_{\bk}F^{\perp\bk}=0$ and $\phi_{\bk}=0$.
 We have not pursued these cases for $D>4$,  they are left for future work.}
   it determines $\bu_\phi$.

The unknown ``velocity'' multipliers $u^{ij}$ can be split into components along and orthogonal to $\phi_{\bk}$:
\bea
u_{\perp\bl}&=&u_\perp \phi_{\bl}+{\hat u}_{\perp\bl},
\qquad \qquad \qquad \hat u_{\perp\bl}\phi^\bl=0,                   \\
u^{\bm\bn}&=&(u^{\bm}\phi^{\bn}-u^{\bn}\phi^{\bm})+\hat u^{\bm\bn}, \quad u^\bm\phi_\bm=0, \quad \hat u^{\bm\bn}\phi_{\bn}=0,
\eea
having, respectively, $1+(D-2)=D-1$ and $(D-2)+(D-2)(D-3)/2=(D-1)(D-2)/2$ components. Using this splitting and \eqref{uphi}, the remaining part of \eqref{two} and \eqref{three} can be, respectively, rearranged into
\be
(\phi_{\bm}F^{\perp\bm}Z_{\bk}\phi^{\bk})u^{\bj}\equiv -(\phi_{\bl}F^{\perp\bl})Z_{\bk}u^{\bk\bj} \approx
\phi_{\bm}\bigl[F^{\perp\bm}X^{\perp\bj}
                -F^{\perp\bj}X^{\perp\bm}\bigr],             \label{uj}
\ee
\be
(\phi_{\bm}F^{\perp\bm})Z_{\bk}\d^{\bk\bl}_{\bi\bj}{\hat u}_{\perp\bl} \approx
\phi_{\bm}\bigl[F^{\perp\bm}X_{\bi\bj}-F_{\bi\bj}X^{\perp\bm}\bigr]. \label{uperpl}
\ee
Generically, \eqref{uj} is $D-2$ equations which can be solved for the $D-2$ components of $u^{\bj}$. Contracting \eqref{uperpl} with $\phi^{\bi}$ yields
\be
(\phi_{\bm}F^{\perp\bm})(\phi^{\bi}Z_{\bi}){\hat u}_{\perp\bj}\approx
\phi_{\bm}\phi^{\bi}\bigl[F^{\perp\bm}X_{\bi\bj}-F_{\bi\bj}X^{\perp\bm}\bigr],
\ee
which (generically) can be solved for the $D-2$ components of ${\hat u}_{\perp\bj}$. The remaining components of \eqref{uperpl} orthogonal to $\phi_{\bk}$ are $(D-1)(D-2)/2-(D-2)=(D-2)(D-3)/2$ \emph{secondary} constraints,
\be
\bar \chi_{\br\bs}:=\d^{\bk\bi\bj}_{\br\bs\bl}\phi^\bl \phi_\bk \phi_\bm \bigl[F^{\perp\bm}X_{\bi\bj}-F_{\bi\bj}X^{\perp\bm}\bigr]. \label{secondary}
\ee
Note the appearance of the projection operator
\be P^{\bi\bj}_{\br\bs}:=\d^{\bk\bi\bj}_{\br\bs\bl}\phi^\bl \phi_\bk,\ee
which projects antisymmetric quantities (multiplied by a factor of $\phi_{\bk}\phi^{\bk}$) onto the subspace orthogonal to $\phi_{\bk}$.

The preservation of the secondary constraint \eqref{secondary} will, upon introducing the values of the known quantities, yield a relation linear in the as-yet-undetermined $u_\perp$, $\hat u^{\bar u\bar v}$:
\bea
0=-\pd_0\chi'_{\br\bs}=\{H_T,\chi'_{\br\bs}\}&\approx&
   \frac12 u^{ij}\{C_{ij},\chi'_{\br\bs}\}+\text{known terms}    \nn \\
&=& u^{\perp\bj}\{C_{\perp\bj},\chi'_{\br\bs}\}
   +\frac12 u^{\bi\bj}\{C_{\bi\bj},\chi'_{\br\bs}\}+\text{known terms}\nn\\
&=&u^{\perp}\phi^{\bj} \{C_{\perp\bj},\chi'_{\br\bs}\}
+\frac14 (\phi_{\bk}\phi^{\bk})^{-1}P^{\bi\bj}_{\bar u\bar v}
\hat u^{\bar u\bar v} \{C_{\bi\bj},\chi'_{\br\bs}\}+\text{known terms}\,,\nn\\
\eea
where $r,s,u,v$ effectively range over the directions orthogonal to $\phi_{\bk}$.

A similar splitting of \eqref{one} gives
\bea
0&\approx& (\phi_{\bk}\phi^{\bk})\bigl[F^{\orth \bj}u_{\orth \bj}
  -\frac12F_{\bi\bj}u^{\bi\bj}+NF^\phi\bigr]                    \nn\\
&=&(\phi_{\bk}\phi^{\bk})F^{\orth \bj}u_\perp
  \phi_\bj-\frac14F_{\bi\bj}P^{\bi\bj}_{\bar u\bar v}\hat u^{\bar u\bar v}
                                         +\hbox{known terms}. \label{chidot}
\eea
Rearranging, this gives $u_\perp$ from
\be
\phi_{\bj}F^{\orth \bj}u_\perp=(\phi_{\bk}\phi^{\bk})^{-1}\frac14
 F_{\bi\bj}P^{\bi\bj}_{\bar u\bar v}\hat u^{\bar u\bar v}+\hbox{known terms}.
\ee
Inserting this into \eqref{chidot} leads to $(D-2)(D-3)/2$ linear relations for the remaining $(D-2)(D-3)/2$ unknowns $\hat u^{\bar u\bar v}$,
\be
\hat u^{\bar u\bar v}P^{\bi\bj}_{\bar u\bar v} \left[ F^{\perp}{}_{\bk}\phi^{\bk}\{C_{\bi\bj},\chi'_{\br\bs}\}
-F_{\bi\bj}\phi^{\bk}\{C_{\perp\bk},\chi'_{\br\bs}\}\right]=\text{known terms}.
\ee

This equation for the remaining undetermined multipliers has the form
\be
\hat u^{\bar u\bar v}(x)D_{\bar u\bar v\br\bs}(x,x')=G_{\br\bs}(x')\,.                                                \lab{lastmultipliereq}
\ee
When one calculates the Poisson brackets $\{C_{ij},\chi'_{\br\bs}\}$, one will get, in general, both terms proportional to the $\d$ function and to its derivative. In view of the derivatives of the $\d$ function buried in $D(x,x')$, it is important to be mindful of the implicit integration over the variable $x$.  Carrying out the integrations by parts (and then, in the end, dropping the prime) will lead to a relation of the form
\be
A^\g_{\bar u\bar v\br\bs} \pd_\g {\hat u}^{\bar u\bar v}+\a_{\bar u\bar v\br\bs} {\hat u}^{\bar u\bar v}=G_{\br\bs}\,.                                              \label{lastmultiplierde}
\ee
Thus, we get \emph{generically} \emph{a system of first-order linear differential equations} for the multipliers $\hat u^{\bar u\bar v}$, the solutions to such a system will thus have a certain degree of nonlocality, in comparison with the solutions of algebraic equations.
The explicit functional forms for $A^\g_{\bar u\bar v\br\bs}$ and $\a_{\bar u\bar v\br\bs}$ in \eqref{lastmultiplierde} can be straightforwardly obtained from the
explicit form of $\{C_{ij},\chi'_{\br\bs}\}$.

Several scenarios are possible. One can determine all the ``missing'' multipliers if this linear relation determines the $\hat u^{\bar u\bar v}$.
Otherwise, some components of this relation may give some additional constraints, which should then be preserved. The chain of constraints could, in principle, go on for several steps before terminating. We cannot exclude the possibility that, in the end, some components of $\hat u^{\bar u\bar v}$ may remain undetermined, so that the solutions have some gauge freedom.  However, we think that these possibilities are quite unlikely.

Generically, the constraints $\pi_\phi$, $C_{ij}$, $\bar\chi_{\br\bs}$ are $1+D(D-1)/2+(D-2)(D-3)/2=(D-1)(D-2)+2$ \emph{second class constraints},
and the number of d.o.f. in the $\phi$-Lorentz sector is
\be
D(D-1)/2+1-\frac12[(D-1)(D-2)+2]=D-1.
\ee
For $D=4$, this gives 3 d.o.f. beyond the metric.
This is what we found in the main text, and exactly agrees with the claim of \cite{LMM11}. For $D>4$ also, the analysis presented here leads to the same number of constraints as presented in that work, however, the formulas and the analysis appearing here are more detailed and simpler.

Although the relations presented here seem  much more tractable than those in \cite{LMM11}, explicitly verifying that the $\chi_{\br\bs}$ are truly second class and their preservation leads to all the missing multipliers is not so easy.  So we cannot yet exclude other possibilities, including the unlikely extreme case that the $\chi_{\br\bs}$ are identically preserved. Then, they would be first class and $(D-2)(D-3)/2$ of the $C_{ij}$ would also be first class.  In this case, the Lorentz sector would have $(D-2)(D-3)$ first class constraints and $1+1+2(D-2)=2(D-1)$ second class constraints.  There are other unlikely possibilities. In any case, we can be sure that there are at least $2(D-1)$ second class constraints and not $D(D-1)/2-1$ first class, unlike the claims of \cite{Ferraro:2018tpu}.

Furthermore, there are indeed (as we had conjectured \cite{Chen1998}) some possibilities for problematical nonlinear constraint effects.
Fixing the multipliers in the second class case requires $\phi_{\bk}F^{\perp\bk}\ne0$ and a nondegeneracy of $D_{\bar u\bar v\br\bs}$.
The dynamics is prone to catastrophic behavior if these quantities degenerate somewhere.  However, if $\phi$ is nonconstant and yet vanishes asymptotically at infinity, \emph{it must have critical points somewhere}, so $\phi_{\bk}$ can be expected to vanish at some points. Thus, indeed, there is good reason to be concerned about the effects of the changing of the rank of the constraint Poisson bracket matrix.

\section{Calculation of \mb{\{C_{ij},\chi'\}}}\label{appG}
\setcounter{equation}{0}

We shall focus here on the part $\{C_{ij},\chi'\}$ of the complete expression $-\pd_0\chi$. The calculation will be organized in several simple tasks.
Start by rewriting $C_{ij}$ in the form
\be
C_{ij}=\cH_{ij}+a_0\phi B_{ij}\, ,\qquad
\cH_{ij}=\pi_{i\bj}-\pi_{j\bi}\,,\qquad B_{ij}=\pd_\a B^{0\a}_{ij}\,.
\ee

In order to explore the dynamical content of $\chi$, it is suitable to rewrite it in the form:
\bea
&&\chi=a_0b\phi^{-1}w_1\big(w_2+2a_0J\phi w_3\big)\,,               \nn\\
&&w_1=\ve^{\bk\bm\bn}\phi_\bk\phi_\br\phi_\bs\, ,                   \nn\\
&&w_2:=T_{\orth\bm\bn}\hpi^{(\br\bs)}\,,                            \nn\\
&&w_3:=\bV^\br T^\bs{}_{\bm\bn}\,,
\eea
see Section \ref{sec6}.
The factors $f=(b,\phi,J)$ are singled out since $\{C_{ij},f\}=0$, see \eq{H.1}. The indices of $w_n$ can be reconstructed by $w_1\to w_1{}^{\bm\bn}_{\br\bs}$, $w_2\to w_2{}_{\bm\bn}^{\br\bs}$, and similarly for $w_3$.

\prg{Step 1.} We begin by calculating the terms $W_n:=\{C_{ij},w'_n\}$, using the formulas
\bea
W_1&=&\{\cH_{ij},(\phi_\bk\ve^{\bk\bm\bn})'\}(\phi_\br\phi_\bs)'
     +\{\cH_{ij},(\phi_\br\phi_\bs)'\}(\phi_\bk\ve^{\bk\bm\bn})'\,,  \nn\\
W_2&=&\{\cH_{ij},T'_{\orth\bm\bn}\}(\hpi^{(\br\bs)})'
   +\{\cH_{ij},(\hpi^{(\br\bs)})'\}T'_{\orth\bm\bn}
   +a_0\phi\{B_{ij},(\hpi^{(\br\bs)})'\}T'_{\orth\bm\bn}\, ,         \nn\\
W_3&=&\{\cH_{ij},(\bV^\br)'\}(T^\bs{}_{\bm\bn})'
      +\{\cH_{ij},(T^\bs{}_{\bm\bn})'\}(\bV^\br)'\,.
\eea
Explicit results are obtained with the help of Appendix \ref{appH}:
\bsubeq\lab{G.4}
\bea
&&W_1=\phi_\bk\big(\d^n_j\ve_\bi{}^{\bk\bm}
  -\d^m_j\ve_\bi{}^{\bk\bn}\big)\d\cdot\phi_\br\phi_\bs
  +\phi_\bi(g_{jr}\phi_\bs
  +g_{js}\phi_\br)\d\cdot\phi_\bk\ve^{\bk\bm\bn}-(i\lra j)\,,        \\
&&W_2=W_{21}+W_{22}+W_{23}\, ,                                       \\
&&\qquad W_{21}=\{\cH_{ij},T'_{\orth\bm\bn}\}(\hpi^{(\br\bs)})'
  =\big(n_i T_{j\bm\bn}+g_{jm}T_{\orth\bi\bn}
                -g_{jn}T_{\orth\bi\bm}\big)\hpi^{(\br\bs)}\d        \nn\\
&&\hspace{85pt}-\big(\vth_{j\g}\pd'_\b\d-\vth_{j\b}\pd'_\g\d\big)
      \big(n_i e_\bm{^\b}e_\bn{^\g}\hpi^{(\br\bs)}\big)'
                                           -(i\lra j)\,,            \nn\\
&&\qquad W_{22}=\{\cH_{ij},(\hpi^{(\br\bs)})'\}T'_{\orth\bm\bn}
        =(\d_j^{(r}\hpi_\bi{}^{\bs)}+\d_j^{(s}\hpi^{\br)}{_\bi})
          T_{\orth\bm\bn}\d -(i\lra j)\,,                           \nn\\
&&\qquad W_{23}=a_0\phi\{B_{ij},(\hpi^{(\br\bs)})'\}T'_{\orth\bm\bn}
  =a_0\phi\pd_\a\big[\big(B^{0\a}_{ij}g^{r\bs}
  +g^{\br k}B^{0\a}_{ki}\d_j^\bs
  +g^{\br k}B^{0\a}_{jk}\d_i^\bs\big)\d\big]T'_{\orth\bm\bn}\,,     \nn\\
&&W_3=W_{31}+W_{32}\, ,                                              \\
&&\qquad W_{31}=\{\cH_{ij},(\bV^\br)'\}(T^\bs{}_{\bm\bn})'
  = (T_{j\bi}{^\br}+\d_j^r\bV_\bi)T^\bs{}_{\bm\bn}\d
                     -n_iT_{\orth\bj}{}^\br(T^\bs{}_{\bm\bn})\d     \nn\\
&&\hspace{85pt}
   -(\vth_{j\g}\pd'_\b\d-\vth_{j\b}\pd'_\g\d)
          (e_\bi{^\b}e^{\br\g}T^\bs{}_{\bm\bn})'-(i\lra j)\,,       \nn\\
&&\qquad W_{32}=\{\cH_{ij},(T^\bs{}_{\bm\bn})'\}(\bV^\br)'
  =\Big[(g_{jm}T^\bs{}_{\bi\bn}-g_{jn}T^\bs{}_{\bi\bm})
  -n_i\big(\d_j^p n^s+\d_j^s n^p\big)T_{p\bm\bn}\Big]\bV^\br\d      \nn\\
&&\hspace{85pt}  -\d_\bi^\bs(\vth_{j\g}\pd'_\b\d-\vth_{j\b}\pd'_\g\d)
                    (e_\bm{^\b}e_\bn{^\g}\bV^\br)'-(i\lra j)\,.     \nn
\eea
\esubeq

\prg{Step 2.} The PB that we are looking for,
\be
\{C_{ij},\chi'\}=\ub{a_0(b\phi^{-1})'W_1(w'_2+2a_0J'\phi'w'_3)}_{Z_1}
 +\ub{a_0(b\phi^{-1})'w'_1(W_2+2a_0J'\phi'W_3)}_{Z_2
                                         =Z_{21}+Z_{22}}\,,   \lab{G.5}
\ee
can be calculated directly from \eq{G.4}.
The term $W_1$ is proportional to the $\d$ function, whereas $W_2$ and $W_3$ contain both $\d$ and $\pd\d$. Terms with $\pd\d$ can be transformed using the $\d$-function identity \eq{H.1}$_1$.

The first term in \eq{G.5} is given by
\be
Z_{1ij}=2a_0(b\phi^{-1})\phi_\bk
      \Big[\d^n_j\ve_\bi{}^{\bk\bm}\phi_\br\phi_\bs
           +\phi_\bi g_{j(\br}\phi_{\bs)}\ve^{\bk\bm\bn}\Big]
      \Big(T_{\orth\bm\bn}\hpi^{(\br\bs)}
           +2a_0J\phi V^\br T^\bs{}_{\bm\bn}\Big)\d-(i\lra j)\,.\lab{G.6}
\ee

The structure of the second term is more complicated, as it contains both $\d$ and $\pd\d$ terms. The contributions to $Z_2(\pd\d)$ are determined by isolating $\pd\d$ terms in $W_2$ and $W_3$:
\bea
W_2(\pd\d)&=&-\Big[(\vth_{j\g}\pd'_\b\d-\vth_{j\b}\pd'_\g\d)
            (n_i e_\bm{}^\b e_\bn{}^\g\hpi^{(\br\bs)})'-(i\lra j)\Big]  \nn\\
&&+a_0\phi\Big[\big(B^{0\a}_{ij}g^{r\bs}
  +g^{\br k}B^{0\a}_{ki}\d_j^\bs
  +g^{\br k}B^{0\a}_{jk}\d_i^\bs\big)\pd_\a\d\Big]T'_{\orth\bm\bn}\, ,  \nn\\
W_3(\pd\d)&=&-(\vth_{j\g}\pd'_\b\d-\vth_{j\b}\pd'_\g\d)
  \Big(e_\bi{^\b}e^{\br\g}T^\bs{}_{\bm\bn}
      +\d_\bi^\bs e_\bm{}^\b e_\bn{}^\g\bV^\br\Big)'-(i\lra j)\,,
\eea
Now, one can insert these terms in \eq{G.5}, substitute the resulting expression  $Z_2(\pd\d;x,x')$ into Eq. \eq{6.13} for the determinant, rearrange the result with the hep of the $\d$-function identity \eq{H.1}$_1$ and integrate over $d^3 x$ (applying the partial integration where needed). Then, replacing $x'$ by $x$, one obtains the first term in the differential equation \eq{6.17}. The second term in \eq{6.17} is produced by the $\d$ function contributions from both $Z_1$ and $Z_2$.

\section{Technical Appendix}\label{appH}
\setcounter{equation}{0}

The formulas presented in this appendix greatly facilitate the work in the ADM basis; see \cite{PM19}. For any variable $U$, we use the notation $U':=U(x')$.
\bea\lab{H.1}
&&fg'\pd'\d=-f\pd(g\d)\,,                                              \nn\\
&&e_\bk{}^0=0\,,\qquad Ne_\orth{}^0=1\,,\qquad \vth^\bk{}_\a= \vth^k{}_\a\,, \nn\\
&&e_\bk{^\a}\vth^k{_\b}=\d^\a_\b\,,\qquad
  e_\bm{^\a}\vth^k{_\a}=\d^k_m-n_mn^k=:\d^k_\bm\,,                      \nn\\
&&\{\pi_i{^\a},N'\}=N^\a n_i\d\,,\qquad
                            \{\pi_i{^\a},J'\}=-J e_\bi{^\a}\d\,,        \nn\\
&&\{\pi_i{^\a},\vth'\}=-\vth e_i{^\a}\d\,,
  \qquad \{\cH_{ij},\vth'\}=\{\cH_{ij},J'\}=0\,,                        \nn\\
&&\{\cH_{ij},U'_\bk\}=(g_{jk}U_i-g_{ik}U_j)\d  \quad
           \text{for}\quad U_\bk=(n_k,\vth_{k\b},e_\bk{}^\b,\phi_\bk)\,,\nn\\
&&\{\cH_{ij},(\d^k_\br)'\}=-(\d^k_j n_r+g_{jr}n^k)n_i\d-(i\lra j)\,.
\eea
\bea\lab{H.2}
&&\{\cH_{ij},\hpi'_{m\bn}\}
  =(g_{jm}\hpi_{i\bn}-g_{in}\hpi_{m\bj})\d-(i\lra j)\,,             \nn\\
&&\{\cH_{ij},\hpi'_{\bm\bn}\}=(g_{jm}\hpi_{\bi\bn}
                      -g_{in}\hpi_{\bm\bj}) -(i\lra j)\,,           \nn\\
&&\{B_{ij}^{0\a},\pi'_k{^\b}\}=2\ve^{0\a\b\g}_{ijkn}\vth^n{_\g}\d
  =(B^{0\a}_{ij}e_k{^\b}+B^{0\a}_{ki}e_j{^\b}
                        +B^{0\a}_{jk}e_i{^\b})\d\,,                 \nn\\
&&\{B_{ij},\hpi'_{k\bn}\}=\pd_\a\big[\big(B^{0\a}_{ij}g_{k\bn}
     +B^{0\a}_{ki}g_{j\bn}+B^{0\a}_{jk}g_{i\bn}\big)\d\big]\,.
\eea
\bea\lab{H.3}
&&\{\cH_{ij},T'_{k\bm\bn}\}=\Big[
       (g_{jm}T_{k\bi\bn}-g_{jn}T_{k\bi\bm})\d
       -g_{ik}(\vth_{j\g}\pd'_\b\d-\vth_{j\b}\pd'_\g\d)
                     (e_\bm{^\b}e_\bn{^\g})'\Big] -(i\lra j)\,,       \nn\\
&&\{\cH_{ij},T'_{\orth\bm\bn}\}=\{\cH_{ij},n^kT_{k\bm\bn}\}
 =(\d^k_j n_i-\d^k_i n_j)T_{k\bm\bn}\d+(n^k)'\{\cH_{ij},T'_{k\bm\bn}\}\nn\\
&&\phantom{x}\hspace{25pt}=\Big[n_i T_{j\bm\bn}\d
          +(g_{jm}T_{\orth\bi\bn}-g_{jn}T_{\orth\bi\bm})\d
          -n'_i(\vth_{j\g}\pd'_\b\d-\vth_{j\b}\pd'_\g\d)
                     (e_\bm{^\b}e_\bn{^\g})'\Big] -(i\lra j),         \nn\\
&&\{\cH_{ij},T'_{\br\bm\bn}\}=\{\cH_{ij},T'_{k\bm\bn}\d^k_\br\}
=\Big[(g_{jm}T_{\br\bi\bn}-g_{jn}T_{\br\bi\bm})\d
   -g'_{i\br}(\vth_{j\g}\pd'_\b\d-\vth_{j\b}\pd'_\g\d)
              (e_\bm{^\b}e_\bn{^\g})'\Big]                            \nn\\
&&\hspace{75pt}
  -n_i\big(\d_j^kn_r+g_{jr}n^k\big)T_{k\bm\bn}\d-(i\lra j)\,,      \nn\\
&&\{\cH_{ij},\bV'_\bn\}=\{\cH_{ij},T_{\bk\bm\bn}\}g^{km}
   =\big(T_{\bj\bi\bn}+g_{jn}\bV_\bi-n_i T_{\orth\bj\bn}\big)\d    \nn\\
&&\hspace{5.5cm} -(\vth_{j\g}\pd'_\b\d-\vth_{j\b}\pd'_\g\d)
                              (e_\bi{^\b}e_\bn{^\g})' -(i\lra j)\,.\quad
\eea

Additional formulas are used in Appendix \ref{appG} to calculate $W_1$ and $W_2$.
\bea
&&\ve^{\a\b\g}\vth^k{_\a}\vth^m{_\b}\vth^n{_\g}=J\ve^{\bk\bm\bn}\,,            \nn\\
&&\{\cH_{ij},(\ve^{\bk\bm\bn})'\}
  =\Big[\d^k_j(\ve_\bi{}^{\bm\bn})'+\d^m_j(\ve^\bk{}_\bi{}^\bn)'
                  +\d^n_j(\ve^{\bk\bm}{}_\bi)'\Big]\d -(i\lra j)\,,  \nn\\
&&\{\cH_{ij},(\phi_\bk\ve^{\bk\bm\bn})'\}
  =\phi'_\bk\Big[\d^n_j(\ve_\bi{}^{\bk\bm})'
             +\d^m_j(\ve_\bi{}^{\bn\bk})'\Big]\d -(i\lra j)\,,       \nn\\
&&\{\cH_{ij},(\phi_\br\phi_\bs)'\}
  =\phi'_\bi(g_{jr}\phi_\bs+g_{js}\phi_{\br})'\d-(i\lra j)\,.
\eea

\section*{Acknowledgments}

We thank Yen Chin Ong and Martin Kr\v{s}\v{s}\'ak for their encouragement and suggestions, and P. Mitri\'c for his help in the early stages of this work.
We also thank the referee for his/her suggestion to explain the breaking of local Lorentz invariance in more detail, and to  Maria Jos\'e Guzman and Alexey Golovnev who's questions led us to make the small but important changes in versions 3 and 4 of this manuscript. One of us (MB) acknowledges the partial support from the Ministry of Education, Science and Technological development of the Republic of Serbia.


\end{document}